\documentclass[12pt,onecolumn,aps,floatfix,showpacs,superscriptaddress]{revtex4}
\usepackage{graphicx}
\usepackage{amsfonts}
\usepackage[T1]{fontenc}
\usepackage[latin9]{inputenc}
\usepackage{amstext}
\usepackage{amssymb}
\usepackage{setspace}

\usepackage{epstopdf,epsfig}
\usepackage{amsmath}

\usepackage{times}
\usepackage{subfigure}
\begin{document}

\title{Fixed Points Structure \& Effective Fractional Dimension \\
for $O(N)$ Models with Long--Range Interactions\vspace{.5cm}}
\author{Nicol\'o Defenu}
\affiliation{SISSA, Via Bonomea 265, I-34136 Trieste, Italy\vspace{.1cm}}
\affiliation{CNR-IOM DEMOCRITOS Simulation Center, Via Bonomea 265, 
I-34136 Trieste, Italy\vspace{.1cm}}
\author{Andrea Trombettoni}
\affiliation{CNR-IOM DEMOCRITOS Simulation Center, Via Bonomea 265, 
I-34136 Trieste, Italy\vspace{.1cm}}
\affiliation{SISSA, Via Bonomea 265, I-34136 Trieste, Italy\vspace{.1cm}}
\affiliation{INFN, Sezione di Trieste, Via Bonomea 265, I-34136 
Trieste, Italy\vspace{.1cm}}
\author{Alessandro Codello\vspace{.25cm}}
\affiliation{$CP^3$-Origins $\&$ the Danish Institute for Advanced Study DIAS,\\
University of Southern Denmark, Campusvej 55, DK-5230 Odense M, Denmark\vspace{1cm}}
\pacs{
11.10.Hi, 05.70.Fh, 11.10.Kk \quad\\
Preprint: CP$^3$-Origins-2014-22 DNRF90 and DIAS-2014-22}

\preprint{}
%\date{\today}

{\setstretch{1.1}
\begin{abstract}
We study $O(N)$ models with power--law interactions 
by renormalization group (RG) methods: 
when the wave function renormalization is not present or not
field dependent, their critical exponents can be computed 
from the ones of the corresponding short--range $O(N)$ models at an effective
fractional dimension.  Explicit results in $2$ and $3$ dimensions are given for the exponent $\nu$. We propose an improved RG 
to describe the full theory space of the models 
where both short--range and long--range interactions are present and competing, and no {\em a priori} choice among the two in the RG flow is done: 
the eigenvalue spectrum
of the full theory for all possible fixed points is drawn and the effective dimension shown to be only approximate. A full description 
of the fixed points structure is given, including multicritical long--range universality classes. 
 \end{abstract}}

\maketitle

{\setstretch{1.3}
$O(N)$ models are celebrated and tireless workhorses of statistical mechanics 
and play a key role in the field of critical phenomena: from one side 
the interest for their properties motivated the developments 
of numerous - analytical and numerical - techniques, from the other side 
they are concretely used as a test ground to benchmark 
the validity of new techniques for critical phenomena and lattice models. 

Among the interactions studied in the context of $O(N)$ models  
an important and paradigmatic role is played by 
long--range (LR) interactions, having the form 
of power--law decaying couplings. A first reason 
is that the results can be contrasted with the findings obtained for 
short--range (SR) interactions, to explore how universal and 
non--universal quantities change increasing the range of the interactions.  
%interestingly, at the critical point the system is 
%scale-invariant, but expected to be not conformal-invariant  
Apart from this motivation {\em per se}, internal to $O(N)$ models, another 
even more important reason for such studies is 
given by the long--lasting interest in understanding the properties 
of systems with LR interactions motivated by their crucial presence
in many systems ranging from plasma physics to astrophysics 
and cosmology \cite{Dauxois10}.
 %
%Dating back to the early days of the renormalization group developments, 
%the study of $O(N)$ models with non-local, eventually LR and power--law, 
%interactions was the subject of intense investigations, starting from 
%the Ising model in one dimension \cite{Dyson69,Thouless69,Anderson70}.\\ 
For a general $O(N)$ model with power--law interactions 
the Hamiltonian reads
\begin{equation}
\label{NvectorSystemHamiltonian}
H=-\frac{J}{2}\sum_{i \neq j} 
\frac{\mathbf{S}_{i}\cdot\mathbf{S}_{j}}{|i-j|^{d+\sigma}}\,,
\end{equation}
where $\mathbf{S}_{i}$ denote a unit vector with $N$ components in the site 
$i$ of a lattice in dimension $d$, $J$ is a coupling energy 
and $d+\sigma$ is the exponent 
of the power--law decay (we refer in the following to cubic lattices).
When $\sigma \leq 0$ a diverging energy density is obtained and to 
well define the thermodynamic
limit it is necessary to rescale the coupling constant $J$ \cite{Campa09}. When $\sigma>0$ the model may have a phase 
transition of the second order, in particular as a function of the 
parameter $\sigma$ three different regimes occur \cite{Fisher72,Sak73}: 
{\em (i)} for $\sigma\leq d/2$ the mean--field approximation is valid even 
at the critical point; 
{\em (ii)} for $\sigma$ greater than a critical value, $\sigma_*$, 
the model has the same critical behaviour of the SR model (formally, 
the SR model is obtained in the limit $\sigma \to \infty$); 
{\em (iii)} for $d/2<\sigma \le \sigma_*$ the system exhibits 
peculiar LR critical exponents. For the Ising model in $d=1$ 
\cite{Dyson69,Thouless69,Anderson70}
the value $\sigma_*=1$ is found, and for $\sigma=\sigma_*$ 
a phase transition of the Berezinskii-Kosterlitz-Thouless universality 
class occur \cite{Cardy81,Froelich82,Luijten01} (see more references in 
\cite{Luijten97}).
Many efforts have been devoted to the determination of $\sigma_*$ and 
to the characterization of the universality classes 
in the region $d/2<\sigma\leq\sigma_*$ for general 
$N$ in dimension $d \ge 2$, which is the case we are going to consider 
in this paper. 
In the classical paper \cite{Fisher72} the expression 
$\eta=2-\sigma$ was found 
for the critical exponent $\eta$ by an $\epsilon$-expansion (at order 
$\epsilon^2$) and 
conjectured to be exact, implying a 
discontinuity in $\sigma_*$, where $\sigma_*=2$ \cite{Fisher72}. \\
A way out was proposed by Sak \cite{Sak73}, who found $\eta=2-\sigma$
for all $\sigma<\sigma_*$ and gave $\sigma_*=2-\eta_{SR}$ (where 
$\eta_{SR}$ is the $\eta$ exponent of the SR model). This $\eta$ is 
a continuous function of $\sigma$ and
there is no correction to the canonical dimension of the field in the case
of LR interactions. Subsequent Monte Carlo (MC) results, based on MC 
algorithms specific for LR interactions \cite{Luijten95}, 
confirmed this picture \cite{Luijten02}.
However the Sak scenario was recently challenged by new 
MC results \cite{Picco12}, suggesting that the 
behavior of the anomalous dimension may be far more complicated that 
the one provided by Sak \cite{Sak73}. Defining the critical exponent 
$\eta_{LR}$ of the $O(N)$ LR models in dimension $d$ with power--law exponent 
$d+\sigma$ as
\begin{equation}
\label{AnomalousDimensionCorrection}
\eta_{LR}(d,\sigma) \equiv 2-\sigma + \delta\eta\,,
\end{equation}
in \cite{Picco12} it was reported that there is a non--vanishing correction 
$\delta\eta$ to Sak's result $\eta=2-\sigma$ in the region 
$d/2<\sigma<\sigma_*$ and that $\sigma_*=2$, as in the earliest work of 
Fisher, Ma and Nickel \cite{Fisher72}. 
In a subsequent work \cite{Blanchard13} 
the presence of a $\delta\eta \neq 0$ was discussed using an 
$\epsilon$--expansion, and as a result the correction $\delta\eta$
should be less than the anomalous dimension of a SR system in dimension $D_{\rm eff}^{BPR}\equiv4+d-2\sigma$ (we refer to such dimension as 
$D_{\rm eff}^{BPR}$ from the authors of \cite{Blanchard13}). 
In the following we are going to show that most of the critical properties
of a LR model in dimension $d$ with power law exponent $d+\sigma$ 
can be inferred from those of a
SR model in the effective fractional dimension $D_{\rm eff}=2d/\sigma \neq 
D_{\rm eff}^{BPR}$, this result being exact in the $N\to\infty$ limit. 
We also observe that the MC results recently presented for a percolation model 
with LR probabilities \cite{Grassberger13} 
seem to agree with the findings of \cite{Picco12} 
and not with the Sak scenario.
In a very recent work new MC results for the Ising model 
with LR interaction in $d=2$ were presented \cite{Parisi14}: 
these results evidence the presence 
of logarithmic corrections into the correlation function
of this kind of systems 
when the value of $\sigma$ is very close to $\sigma=2-\eta_{SR}$, 
implying the numerical difficulty of extracting 
reliable results for the critical exponents with small error bars around 
$\sigma=2-\eta_{SR}$.

The controversy about the actual value of $\sigma_*$ raised by 
recent MC results has not really a compelling 
quantitative {\em raison d'\^{e}tre}: 
after all, for the Ising model in $d=2$ it is $\eta_{SR}=1/4$ and 
$\sigma_*=7/4$ predicted by Sak should be contrasted with 
$\sigma_*=2$ suggested in \cite{Picco12} 
(even though 
the value of $\eta$ at $\sigma=7/4$ obtained in \cite{Picco12} is 
$\eta = 0.332$ and it should be contrasted with $\eta=1/4$ predicted 
by Sak). The issue raised by recent MC results is rather of principle, since 
it generally questions how the LR terms ($p^\sigma$) renormalize and especially 
how the SR term ($p^2$) in the propagator is dressed by the presence of LR 
interactions.
In this paper we aim at clarifying such issues using a 
functional renormalization group approach \cite{Berges02,Delamotte07}. 
%The advantages of the method are that it allows for to treat on equal foot 
%general values of $N$, $d$ and $\sigma$;
%it allows a simple way to compute the critical exponent $\nu$; and 
%can be used to study efficiently multicritical fixed points \cite{Codello13}.

%\subsection*{
%FRG approach
%Renormalization group analysis
%}
%
We are interested in universal quantities, and as usual 
we replace the spin variables $\{\mathbf{S}_{i}\}$ 
with an $N$--component vector field $\boldsymbol{\phi}(x)$ in continuous space. 
We define a scale dependent effective action $\Gamma_{k}$ 
depending on an infrared cutoff $k$ and on the continuous field 
$\boldsymbol{\phi}$: 
when $k\rightarrow k_0$, where $k_0$ is some ultraviolet scale, the effective action is equal to the mean--field 
free energy of the system, while for $k\rightarrow 0$ it 
is equal to the exact free energy \cite{Berges02}.
Our first ansatz for the effective action reads
\begin{equation}
\label{EffectiveAction}
\Gamma_{k}[\phi]=\int d^{d}x \left\{Z_{k}\partial^{\frac{\sigma}{2}}_{\mu}
\phi_{i}\partial^{\frac{\sigma}{2}}_{\mu}\phi_{i}+U_{k}(\rho)\right\}\,,
\end{equation}
where the summation over repeated indexes is assumed,
$\rho=\frac{1}{2}\phi_{i}\phi_{i}$, and
$\phi_{i}$ is the $i$--th component of $\boldsymbol{\phi}$.
The notation $\partial^{\frac{\sigma}{2}}_{\mu}$ is a compact way to intend that
the inverse propagator of the effective action \eqref{EffectiveAction} 
in Fourier space depends on $q^{\sigma}$ and not on $q^{2}$ as in the SR case.
$Z_{k}$ is the wave function renormalization of the model that at this 
level of approximation is field independent.
The effective potential $U_{k}(\rho)$ satisfies 
a renormalization group equation \cite{Wetterich93}; when this is
rewritten in terms of dimensionless variables (denoted by bars) one can find 
the fixed points, or scaling solutions, $\bar{U}_{*}(\bar{\rho})$ 
by solving it \cite{Codello12}.
Using an infrared cutoff suited for LR interactions,
$R_{k}(q)= Z_k (k^{\sigma}-q^{\sigma})\theta(k^{\sigma}-q^{\sigma})$,
we obtain the flow equation for the effective potential,
\begin{equation}
\begin{split}
&\partial_t \bar{U}_{k}= -d\bar{U}_{k}(\bar{\rho})+(d-\sigma+\delta\eta)\bar{\rho}\,\bar{U}'_{k}(\bar{\rho})\\
&+\frac{\sigma}{2} c_d (N-1)\frac{1-\frac{\delta\eta}{d+\sigma}}{1+\bar{U}'_{k}(\bar{\rho})}
+\frac{\sigma}{2} c_d \frac{1-\frac{\delta\eta}{d+\sigma}}{1+\bar{U}'_{k}(\bar{\rho})+2\bar{\rho}\,\bar{U}''_{k}(\bar{\rho})}\,,
\label{EffectivePotentialLPAprimeEquation}
\end{split}
\end{equation}
where $c_d^{-1}=(4\pi)^{d/2} \Gamma\left(d/2+1 \right)$ 
and $\delta\eta$ is an eventual anomalous dimension correction, 
related to the flow of the wave function renormalization by $\delta \eta = - \partial_t \log Z_k$, where $t=\log(k/k_0)$ is the RG time 
and $k_0$ is the ultraviolet scale.
%\subsection*{
%LPA approximation
%Effective dimension}

We start our analysis 
considering the case $Z_{k}=1$, which implies $\delta\eta=0$. 
%From Eq.\eqref{EffectivePotentialLPAprimeEquation}
It is then possible to show that
%, for an $O(N)$ LR model in dimension $d$ and 
%power--law exponent $d+\sigma$,
the flow equation \eqref{EffectivePotentialLPAprimeEquation} for the effective 
potential can be put in relation with the corresponding equation 
for a SR model \cite{Codello12,Morris94} in an effective fractional dimension
\begin{equation}
\label{DimensionaEquivalence}
D_{\rm eff}=\frac{2d}{\sigma}
\end{equation}
(in the following we denote by capital 
$D$ the dimension of the SR $O(N)$ model). 
Namely, we can see that the LR and SR universality classes 
in, respectively, dimension $d$ and $D_{\rm eff}$ 
are the same at this level of approximation.
%as can be seen comparing flow equation \eqref{EffectivePotentialLPAprimeEquation},
%in the $\delta\eta=0$ limit, with
%its SR equivalent \cite{Codello12,Morris94}.
%
The equivalence between the fixed point structure of these
two models can also be seen using the spike plot technique
described in  \cite{Codello12,Morris94}: the corresponding figure
may be found in the Appendix A. 
%Since we are considering LR interactions the Mermin--Wagner 
%theorem \cite{Codello13} does not apply, and also 
%systems with continuous symmetries, such as the $N\geq2$ models, 
%can undergo continuous phase transitions in $d=2$.
%
%\begin{figure}[ht]
%\centering
%\label{Long_Range_Spike_Plot_N=1}
%\includegraphics[scale=0.7]{FIGURE1.pdf}
%\caption{Each value of $\Sigma \equiv \bar{U}_{*}'(0)$ 
%for which we have a spike in the above figure is the derivative at the origin 
%of a well defined fixed point effective potential: thus every spike 
%is the signature of a different universality class. 
%Solid lines represents spike plots of LR models 
%in dimension $d$ with power--law exponent $\sigma$, 
%while dashed lines represent spike plots of 
%SR models in dimension $D=D_{\rm eff}=2d/\sigma$.
%The plot is for the case $N=1$ and $d=2$ for the cases 
%$\sigma = 1.25,1.75,1.9$.
%}
%\label{Fig1}
%\end{figure}

From this analysis it follows that  
by varying $\sigma$ at fixed $d$ we go trough a sequence of $\sigma_{c,i}$ 
at which new multicritical LR universality classes appear,
%In particular, these are lower critical parameters above which new scaling solutions appear,
in a way analogous to the sequence of upper critical dimensions found
in SR models as $d$ is varied \cite{Codello12}.
For the Ising universality class 
the lower critical decay exponent is $\sigma_{c,2}=d/2$ 
in agreement with known results \cite{Fisher72}. 
In the case of a $i$--th multicritical model with LR interaction 
the lower critical decay exponent is found to be 
$\sigma_{c,i}=\frac{d(i-1)}{i}$.
Since the new fixed points branch from the Gaussian 
fixed point, their analysis based on the
ansatz \eqref{EffectiveAction}, first term of an expansion of the effective action in powers of the anomalous dimension, is consistent
%
%the correspondence between universality classes of SR models in dimension $D_{\rm eff}$ and
%LR models in dimension $d$ with decay exponent $\sigma$,
and the existence of multicritical LR $O(N)$ models can be extrapolated 
to be valid in the full theory.

Within this approximation it is also possible to establish a mapping
between the LR correlation length exponent $\nu_{LR}(d,\sigma)$ and the
equivalent SR one $\nu_{SR}(D_{\rm eff})$. The relation is found to be:
\begin{equation}
\label{NuRelationLPA}
\nu_{LR}(d,\sigma)=\frac{2}{\sigma}\nu_{SR}(D_{\rm eff})\,.
\end{equation}
As a check, we observe that relations 
\eqref{DimensionaEquivalence} and \eqref{NuRelationLPA} 
are satisfied exactly by the spherical model 
\cite{Joyce}. 
In fact in the $N \to \infty$ limit 
our approximation provides exact critical exponents \cite{Tetradis94}.
\begin{figure}
\centering
\includegraphics[scale=.85]{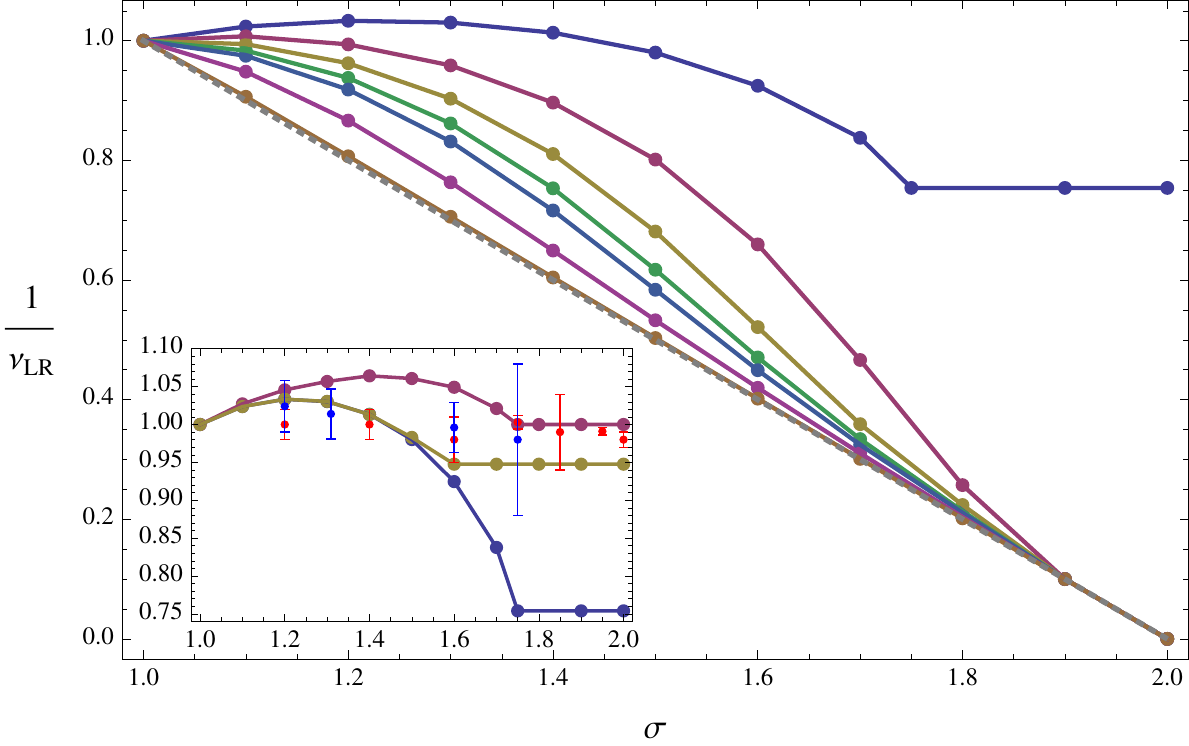}
\caption{
$y_{t}=1/\nu_{LR}$ exponent as a function of $\sigma$ in $d=2$ for some values of $N$ (from top: $N=1,2,3,4,5,10,100$).
The dashed line is the analytical result obtained for the spherical model $N=\infty$.
Inset: $y_{t}=1/\nu_{LR}$ vs. $\sigma$ for the $d=2$ LR Ising model compred
with MC data of \cite{Luijten02} (red circles) 
and of \cite{Parisi14} (blue circles). 
The three continuous lines represents the estimates made using \eqref{NuRelationLPA2} with the 
numerical values of $\nu_{SR}(D_{\rm eff}')$ and $\eta_{SR}(D_{\rm eff}')$ taken from
recent high--precision estimates in fractal dimensions 
\cite{El14} (top red line),
from \cite{Codello13,Codello14} where the $O(N)$ model
definition for $\eta_{SR}$ is used (blue bottom line)
and from \cite{Codello12} where the Ising definition of $\eta_{SR}$ is used instead (yellow middle line)
\cite{El14}.
}
\label{3}
\end{figure}
%
%
%\subsection*{
%LPA$'$ approximation
%Anomalous dimension}

To study anomalous dimension effects one has 
to study the equation for the effective potential $U_{k}$   
in the case $\delta\eta\neq 0$, i.e. when $Z_k$ in (\ref{EffectiveAction}) 
is non--constant.
One obtains the scale derivative of the wave function renormalization from
$\partial_tZ_{k}=\lim_{p\rightarrow 0}\frac{d}{dp^{\sigma}} \partial_t\Gamma_{k}^{(2)}(p,-p)$
and computes the anomalous dimension using 
$\delta \eta = - \partial_t \log Z_k$.
Since the flow equation generates no non--analytic terms in $p$, from this definition we find $\delta\eta=0$, 
in agreement with Sak's result \cite{Sak73}, in which the anomalous 
dimension does not get any non mean--field contribution.
However, an anomalous dimension is present, at this approximation level, in 
the SR system, thus we obtain a new dimensional 
equivalence:
\begin{equation}
\label{EffectiveDimensionLPA$'$}
D_{\rm eff}'=\frac{[ 2-\eta_{SR}(D_{\rm eff}') ] d}{\sigma}\,,
\end{equation}
which is in agreement with the results of the dimensional analysis performed 
for the Ising model in \cite{Parisi14} and with the arguments presented 
for the LR and SR Ising spin glasses in \cite{Young12}.
Eq.\,\eqref{EffectiveDimensionLPA$'$} is valid for any $N$ and it is an implicit equation for 
$D_{\rm eff}'$: to find $D_{\rm eff}'$ one has to know the critical exponent $\eta_{SR}$ 
in fractional dimension \cite{Codello13,Katz77,Holovatch93,El14}.
At date the most precise evaluation of $\eta_{SR}$ 
for the Ising model ($N=1$) in fractional dimension is given in \cite{El14};
results for general $N$ are given by in 
\cite{Codello13}, turning in rather good agreement with \cite{El14} for $N=1$ 
and with \cite{Holovatch93} for $N \ge 2$. 
%Using these results we can then evaluate $D_{\rm eff}'$; a plot of it 
%s reported in Appendix \ref{Appendix_A}.

In the case of a running, not field dependent, 
wave function renormalzation we also 
obtain the following relation for the critical exponent $\nu_{LR}$:
\begin{equation}
\label{NuRelationLPA2}
\nu_{LR}(d,\sigma)=\frac{2-\eta_{SR}(D_{\rm eff}')}{\sigma}\, \nu_{SR}(D_{\rm eff}')\,.\\
\end{equation}
In Fig.\,\ref{3} we compare the exact behaviour for the $y_{t}=1/\nu_{LR}$
LR exponent in the spherical $N \to \infty$ limit 
with the behaviour obtained using the effective dimension 
$D'_{eff}$ for various values of $N$.
In the inset of Fig.\,\,\ref{3} we plot MC results from 
\cite{Luijten02} and \cite{Parisi14} 
together with the results obtained by the effective dimension $D_{\rm eff}'$ both 
at our approximation level and with the use of high--precision estimates of the SR critical exponents in fractal dimensions   
from \cite{El14} in \eqref{EffectiveDimensionLPA$'$}.
We expect these results to be more reliable as $N$ grows due
to the relative decrease of anomalous dimensions effects in these cases.
Relations \eqref{EffectiveDimensionLPA$'$} and \eqref{NuRelationLPA2}
can be also used to extend this analysis to multicritical fixed points in LR systems.
We also note the fact that in $d=2$ for every $N\geq2$ the exponent $y_{t}$ goes to zero, and 
thus $\nu_{LR}$ goes to infinity, 
is a consequence of, and consistent with, the Mermin--Wagner theorem \cite{Codello13}.
%
%Even these results, not reported here, are expected to be more accurate that the standard bi--critical ones.
%It is also possible to use the presented results to study other universality classes (tri--critical, and so on) 
%using the effective dimension \eqref{EffectiveDimensionLPA$'$};
%these results, not reported here, are expected to be 
%more accurate than the (standard bi--critical) Ising ones, 
%since the anomalous dimension in SR multicritical models 
%is smaller than in the Ising case and then this approximation
%is more reliable.

In Fig.\,\ref{nudimension3} we plot the exponent $y_{t}$ for various $N$ in three
dimensions using \eqref{EffectiveDimensionLPA$'$}: due to the better performances of our approximation in three dimensions, 
we expect these results to be quantitatively very reliable, when compared with future numerical simulations.
The curves of Fig.\,\ref{3} and Fig.\,\ref{nudimension3} are genuine 
universal predictions of our analysis and to our knowledge are new.
\begin{figure}
\centering
\includegraphics[scale=1]{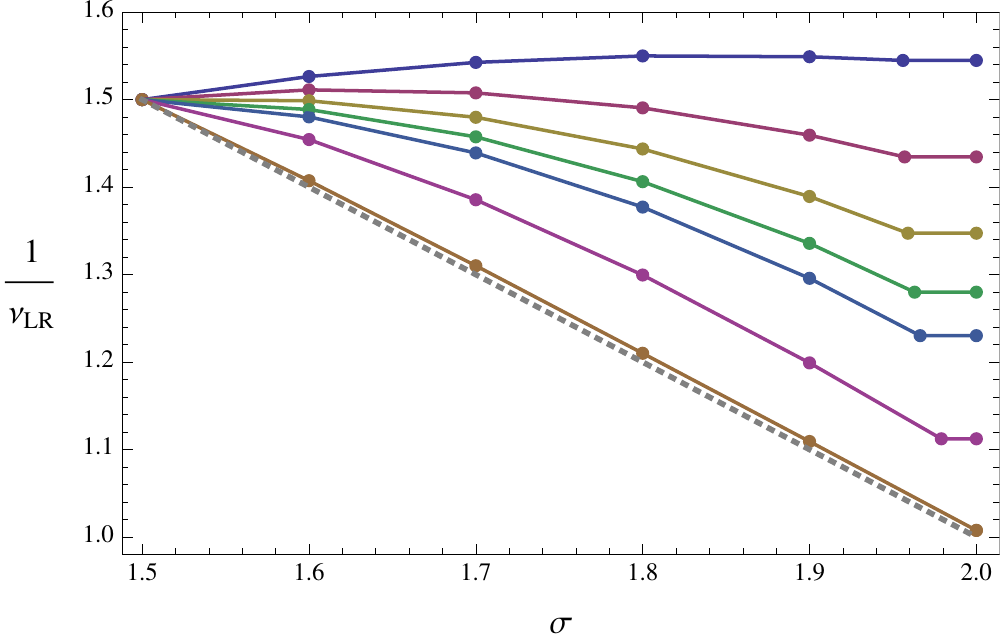}
\caption{$y_{t} = 1 / \nu_{LR}$ exponent as a function of $\sigma$ in $d=3$ for some values of $N$ (from top: $N=1,2,3,4,5,10,100$). As in  Fig. \ref{3} the dashed line is the analytical result obtained for the spherical model.
}
\label{nudimension3}
\end{figure}
%
%
%\subsection*{
%LPA$''$ Approximation
%Competition between SR and LR interactions}

The present analysis suggests 
the validity of Sak's results for the value of $\sigma_*$. 
On the other hand, since the ansatz \eqref{EffectiveAction} 
does not contain any SR term,
such approximation is not able to describe the case $\sigma>\sigma_*$, 
in which SR interactions could become dominant. 
In order to investigate these effects we enlarge our 
theory space and we propose the new ansatz:
\begin{equation}
\label{EffectiveAction_LPA$''$}
\Gamma_{k}[\phi]=\!\int\! d^{d}x\! \left\{\!Z_{\sigma,k}\partial^{\frac{\sigma}{2}}_{\mu}
\phi_{i}\partial^{\frac{\sigma}{2}}_{\mu}\phi_{i}\!+\!Z_{2,k}\partial_{\mu}\phi_{i}\partial_{\mu}\phi_{i}\!+\!U_{k}(\rho)\!\right\}\,,
\end{equation}
where we have both LR and SR terms in the propagator. A similar ansatz was introduced in \cite{Tissier14,Balog14} 
where the dimensional reduction of the Ising model with LR interaction in presence of disorder was studied.
\begin{figure}
\centering
\includegraphics[scale=.7]{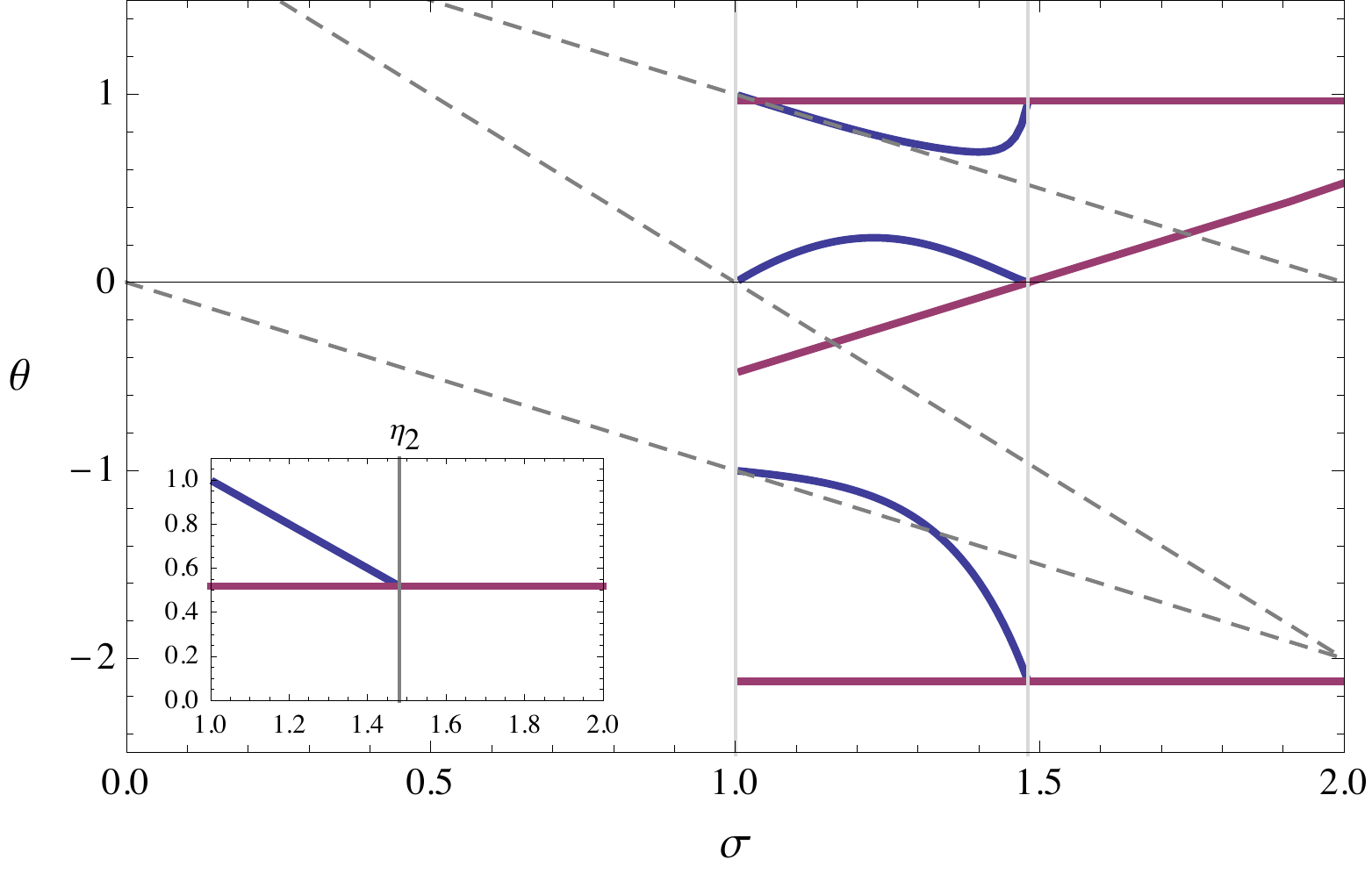}
\caption{Eigenvalues ($\theta$) of the RG stability matrix in $d=2$ as a function 
of $\sigma$ for the SR (red lines) and LR (blue lines) fixed points.
Mean field exponent are represented by dashed lines.
The vertical lines mark $\sigma=\frac{d}{2}=1$ and $\sigma_*=2-\eta_{SR}$.
For $1< \sigma < \sigma_*$ both fixed points are present, 
but the LR one has two IR attractive directions, while the SR has one.
For $\sigma >  \sigma_*$ only the SR fixed point is present, while for $\sigma < 1 $ the LR fixed point is Gaussian and the exponent are mean field.
%Thus for every value of $\bar{J}_{\sigma,0}$ and $\lambda_0$, 
%if $\kappa_k$ is tuned to the specific critical value 
%the RG flow will lead to the LR fixed point and not the the SR one.
Inset:  anomalous dimension $\eta_2$ vs. $\sigma$ in the same case.
}
\label{Figure4}
\end{figure}
%

%As a first step to further proceed we need to define a proper cutoff for 
%this theory; in principle one should cutoff just the 
%dominant term in the propagator, however 
%we do not know a priori which term will be dominant, 
%due to the competition between SR and LR interactions near $\sigma_*$.
%
We need to choose a proper cutoff function for the propagator of 
the theory \eqref{EffectiveAction_LPA$''$}. Since we do not know
{\em a priori} which will be the dominant term for $\sigma\simeq\sigma^{*}$
we take the following combination:
\begin{equation}
\label{Cutoff_Function_LPA$''$}
R_{k}(q)=Z_{\sigma,k} (k^{\sigma}-q^{\sigma})\theta(k^{\sigma}-q^{\sigma})+
Z_{2,k} (k^{2}-q^{2})\theta(k^{2}-q^{2})\,.
\end{equation}
%
%The two possible choices to properly define the dimensionless coupling are 
%discussed in detail in Appendix \ref{Appendix_B}
%
The ansatz (\ref{EffectiveAction_LPA$''$}) and the cutoff choice (\ref{Cutoff_Function_LPA$''$}) are consistent with the ones of the previous analysis when LR interactions are dominant, but they are still valid when SR become important and will allow us to study the
%the results obtained with ansatz \eqref{EffectiveAction_LPA$''$} includes those 
%obtained by ansatz \eqref{EffectivePotentialLPAprimeEquation}, extending them in the 
whole $\sigma$ range.
The general flow equations that follows and further details are reported in the Appendix B.
 
To further proceed, we make a Taylor expansion 
of the effective potential around its minimum 
and we maintain only the lowest terms: 
$\bar{U}_{k}(\bar{\rho})=\frac{1}{2}\lambda_{k}(\bar{\rho}-\kappa_{k})^{2}$. In addition to the equations 
for $\lambda_k$ and $\kappa_k$ 
we have also an equation
for the anomalous dimension $\eta_{2}=-\partial_{t} \log Z_{2,k}$ 
and one for the LR coupling $J_{\sigma,k}\equiv Z_{\sigma,k} / Z_{2,k}$.
Here we report these equations in the two dimensional $N=1$ case:
\begin{eqnarray}
%
%\begin{equation}
\eta_{2} &=& \frac{(2+\sigma\bar{J}_{\sigma,k})^{2}\kappa_{k}\lambda_{k}^{2}}{(1+\bar{J}_{\sigma,k})^{2}(1+\bar{J}_{\sigma,k}+2\kappa_{k}\lambda_{k})^{2}}\nonumber\\
%\end{equation}
%
%\begin{equation}
%\label{opportuna}
\partial_{t}\kappa_{k} &=&  \;   -\eta_2 \kappa_k + 3\frac{1-\frac{\eta_2 }{d+2}+\frac{\sigma}{2}\bar{J}_{\sigma,k}}{(1+\bar{J}_{\sigma,k}+2 \kappa_k  \lambda_k)^2\nonumber}\\
%+(N-1)\frac{1-\frac{\eta_2}{d+2}+\frac{ \sigma}{2}\bar{J}_{\sigma,k}}{(1+\bar{J}_{\sigma,k})^2}\,,
%\end{equation}
%
%\begin{equation}
\partial_{t} \lambda_{k} &=& 2(-1+\eta_{2})\lambda_{k} + 18 \lambda_k\frac{1-\frac{\eta_2 }{d+2}+\frac{\sigma}{2}\bar{J}_{\sigma,k}}{(1+\bar{J}_{\sigma,k}+2 \kappa_k  \lambda_k)^3}\nonumber\\
%+ 2 \lambda_k(N-1)\frac{1-\frac{\eta_2}{d+2}+\frac{ \sigma}{2}\bar{J}_{\sigma,k}}{(1+\bar{J}_{\sigma,k})^3}
%
%\begin{equation}
\partial_{t}\bar{J}_{\sigma,k} &=& (\sigma-2)\bar{J}_{\sigma,k}+\eta_{2}\bar{J}_{\sigma,k}\,.
%\end{equation}
%\end{equation}
%
\label{CouplingSet}
\end{eqnarray}
%
%
%Following the renormalization of this four 
%quantities, as is shown in Appendix \ref{Appendix_B}, we found 
%$\eta = 2-\sigma$ for $\sigma < 2-\eta_{SR}$ and  $\eta=\eta_{SR}$ for 
%$\sigma > 2-\eta_{SR}$ in agreement with the Sak's result 
%$\sigma_*=2-\eta_{SR}$ .
%
Using these equations we are able to describe in detail the structure of the phase diagram.
The anomalous dimension of LR $O(N)$ models is still $\eta=2-\sigma$, then for $\sigma>\sigma_{*}=2-\eta_{SR}$
the dimensionless coupling $\bar{J}_{\sigma,k}$ is always renormalized to zero, 
whatever initial conditions we choose and 
the system behaves as if only SR interactions were present. 
On the other hand when $\sigma<\sigma_*$ 
a new interacting LR fixed point branches from the SR one and is characterized by a finite 
value of $\bar{J}_{\sigma,*}$. 

In Fig.\,\ref{Figure4} we show the critical exponents of both SR and LR fixed 
points obtained from the coupling set \eqref{CouplingSet}.
The SR fixed point has just one repulsive direction for $\sigma>\sigma_*$ (the standard Wilson--Fisher one)
and the LR fixed point does not exist at all. 
At $\sigma=\sigma_*$ the smallest attractive eigenvalue 
hits zero and the LR fixed point emerges from the SR fixed point.
For $\sigma<\sigma_*$, the SR fixed points has two repulsive directions 
while the LR one has just one repulsive direction.
Finally at $\sigma=\frac{d}{2}=1$ the LR fixed point becomes Gaussian and
for all $\sigma<\frac{d}{2}=1$ the behavior is mean field.

From the analysis of Fig.\,\ref{Figure4} one clearly understands that  
the LR fixed point is attractive along the direction which connects it
to the SR (Wilson--Fisher) fixed point, thus for $\sigma<\sigma_*$ the
SR fixed point becomes repulsive in the $\bar{J}_{\sigma,k}$ direction and
the LR fixed point controls the critical properties of the system.
In the $\sigma \rightarrow \sigma_*$ limit the LR fixed point moves towards the 
SR one and finally merges with it at $\sigma = \sigma_*$.
%
%In the inset we show the anomalous dimension: in turns out to be SR above $\sigma_*$ and LR below, exactly as predicted by Sak.
%
This structure for the phase diagram implies that the anomalous dimension is given (as show in the inset of Fig.\,\ref{Figure4})
by the LR value $\eta_2 = 2-\sigma$ for $\sigma < 2-\eta_{SR}$ and
by the SR value $\eta_2 = \eta_{SR}$ for $\sigma > 2-\eta_{SR}$, thus confirming Sak's scenario.
It is important to stress that we are not imposing this picture by hand, but it emerges dynamically form the solution of \eqref{CouplingSet}.
It is also important to underline that the threshold $\sigma^{*}=2-\eta_{SR}$ is also generated dynamically, 
with the SR anomalous dimension appearing in it being the one pertinent to the approximation level considered.
%
%This analysis has also the merit to explicitly present a set of beta functions exhibiting a phase space supporting Sak's scenario.

%\subsection*{Summary and conclusions} 
{\em Conclusions:} We studied $O(N)$ long--range (LR) models 
in dimension $d \ge 2$.
Using the flow equation for the effective potential
alone we found 
that universality classes of $O(N)$ LR models are 
in correspondence with those of $O(N)$ short--range (SR)  
models in effective dimension $D_{\rm eff}=2d/\sigma$.
We also found new multicritical potentials which are present, at fixed $d$, 
above certain critical values of the parameter $\sigma$. 
%As a check, since for large $N$ this approximation is exact, 
%we retrieve an exact result linking the critical exponents 
%of spherical models with LR and SR interactions.
 
We then considered anomalous dimension effects considering
also the flow of a field independent wave function renormalization. 
Extending the approach described in \cite{Delamotte07} to the
LR case we
found $\delta\eta=0$, i.e. the Sak's result \cite{Sak73} 
in which there are no correction to the mean--field value 
of the anomalous dimension. 
The relation between the LR model %of parameters $(d,\sigma)$  
and the SR model is now valid at the effective dimension $D_{\rm eff}'$ 
defined by Eq.\,\eqref{EffectiveDimensionLPA$'$},
while the correlation length exponent is given according 
to Eq.\,\eqref{NuRelationLPA2}. 
Quantitative predictions for the exponent $\nu_{LR}$ for various values of 
$N$ were as well presented in $d=2$ and  $d=3$.

Finally we introduced an effective action
where both the SR and LR terms
are present. This approach does not impose {\it a priori} 
which is the dominant coupling in the RG flow. 
We showed how  Sak's result is again justified 
by the fixed point structure of the model, 
where a LR interacting fixed point appears only 
if $\sigma<\sigma_*$ and controls the critical behavior of the system. 
Interestingly, the effective dimension $D_{\rm eff}'$ can be shown 
not to be exact 
at this approximation level: however it is 
possible to estimate the error committed using the 
effective dimension $D_{\rm eff}'$, this error being 
proportional to the ratio between SR and LR couplings.   

The final picture emerging from the our analysis is the following: starting at 
$\sigma=0$ and increasing $\sigma$ towards $2$ we have that for $\sigma<d/2$ 
only the LR Gaussian fixed  point exists and no SR terms in the propagator are 
present. At $\sigma=d/2$ a new interacting 
fixed point emerges from the LR Gaussian one and the same happens at the 
values  
$\sigma_{c,i}$ where new LR universality classes appear (in the same way as 
the multicritical SR fixed points are generated below the upper critical 
dimensions). Finally, when $\sigma$ approaches $\sigma_*$ the LR Wilson--Fisher 
fixed point merges with its SR equivalent and the LR term in the propagator 
disappears for $\sigma>\sigma_*$: this 
has to be contrasted with the case $\sigma<\sigma_*$ 
where at the interacting LR fixed points the propagator 
contains also a SR term. 
The same scenario is valid for all multicritical 
fixed points, provided that the $\sigma_*$ values  are computed 
with the corresponding SR anomalous dimensions.
\\
\\
\textit{Acknowledgements.} 
We are very grateful to G. Gori and M.A. Rajabpour for many useful 
discussions during various stages of the work.
We also acknowledge useful correspondence with I. Balog, G. Tarjus and M. Tissier.
The CP$^3$-Origins centre is partially funded by the 
Danish National Research Foundation, grant number DNRF90.
\\
\\
\textit{Note added:}
During the final phase of this work a 
paper on LR interactions appeared on the arXiv 
\cite{Parisi14_2}, 
showing that for $\sigma\simeq\sigma_*$ 
logarithmic corrections to the correlation function are present. 
We observe that the vanishing of the smallest attractive eigenvalues 
for the LR fixed point at $\sigma=\sigma_*$ shown in Fig.\ref{Figure4} 
is in agreement with such finding. 
%These unexpected corrections could be responsible for the unexpected behaviour of MC results in \cite{Picco13} and \cite{Grassberger13}. 
%
\appendix
\section{Pure long--range analysis}\label{Appendix_A}
Let us consider a Ising model where the spins interacts
via a long--range (LR) potential, with power low decaying interactions: 
the Hamiltonian reads
\begin{equation}
\label{NvectorSystemHamiltonian}
H=-\frac{J}{2}\sum_{i \neq j} 
\frac{\mathbf{S}_{i}\cdot\mathbf{S}_{j}}{|i-j|^{d+\sigma}}\,,
\end{equation}
where $d$ is the dimension of the model and $d+\sigma$ the exponent
of the power low decaying potential. 
We study the continuous field model analogous to the Hamiltonian
\eqref{NvectorSystemHamiltonian}. The effective action in the pure (LR) case reads:
\begin{equation}
\label{EffectiveAction_app}
\Gamma_{k}[\phi]=\int d^{d}x \left\{Z_{k}\partial^{\frac{\sigma}{2}}_{\mu}
\phi_{i}\partial^{\frac{\sigma}{2}}_{\mu}\phi_{i}+U_{k}(\rho)\right\}\,,
\end{equation}
where the summation over repeated indexes is assumed,
$\rho=\frac{1}{2}\phi_{i}\phi_{i}$ and $\phi_{i}$ is the $i$--th component of $\boldsymbol{\phi}$.
The notation $\partial^{\frac{\sigma}{2}}_{\mu}$ is a compact way to intend that
the inverse propagator of the effective action \eqref{EffectiveAction_app} 
in Fourier space depends on $q^{\sigma}$.
The effective potential $U_{k}(\rho)$ in \eqref{EffectiveAction_app} 
obeys the evolution equation derived in \cite{Wetterich93}. 
The evolution equation for the potential is as usual 
rewritten in terms of dimensionless variables,
\begin{align}
\bar{U}_{k}(\bar{\rho})&=k^{-d}U_{k}(\rho)\,,\\
\bar{\rho}&= Z_{k}k^{\sigma-d}\rho\,,\\
\bar{q}&=k^{-1}q\,,
\end{align}
and then equated to zero, in order to find the fixed point solution 
$\bar{U}_{*}(\bar{\rho})$ \cite{Codello12}.
We define a generalized Litim cutoff suited for long--range (LR) interactions:
\begin{equation}
R_{k}(q)=Z_{k}(k^{\sigma}-q^{\sigma})\theta(k^{\sigma}-q^{\sigma})\,.
\label{GeneralizedLitimCutoff}
\end{equation}
Using \eqref{GeneralizedLitimCutoff} we obtain the equation:
\begin{equation}
\begin{split}
\partial_t \bar{U}_{k}=& -d\bar{U}_{k}(\bar{\rho})+(d-\sigma+\delta\eta)\bar{\rho}\,\bar{U}'_{k}(\bar{\rho})
+\frac{\sigma}{2} c_d (N-1)\frac{1-\frac{\delta\eta}{d+\sigma}}{1+\bar{U}'_{k}(\bar{\rho})}\\
&+\frac{\sigma}{2} c_d \frac{1-\frac{\delta\eta}{d+\sigma}}{1+\bar{U}'_{k}(\bar{\rho})+2\bar{\rho}\,\bar{U}''_{k}(\bar{\rho})}\,,
\label{EffectivePotentialLPAprimeEquation_app}
\end{split}
\end{equation}
where $c_d^{-1}=(4\pi)^{d/2} \Gamma\left(d/2+1 \right)$ and $\delta\eta$ is defined by
\begin{equation}
\label{EtaDefinition_app}
\delta\eta=-\frac{1}{Z_{k}} \partial_t Z_k \,, %\frac{dZ_{k}}{d\log k}\,,
\end{equation}
is an eventual non--mean field correction to 
the anomalous dimension of the model, 
i.e. $\eta_{LR} \equiv 2-\sigma + \delta\eta$.
Here $t=\log (k/ k_0)$ is the RG time.
\begin{figure}[ht]
\centering
%\subfigure[]{
\label{Long_Range_Spike_Plot_N=2}
\includegraphics[scale=0.7]{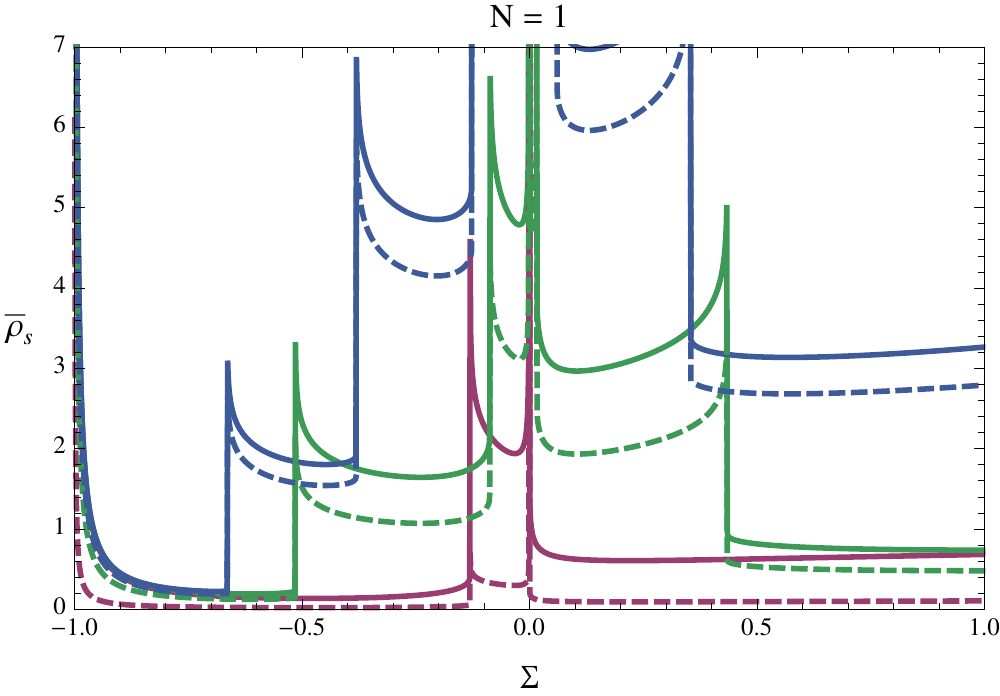}\qquad\includegraphics[scale=0.7]{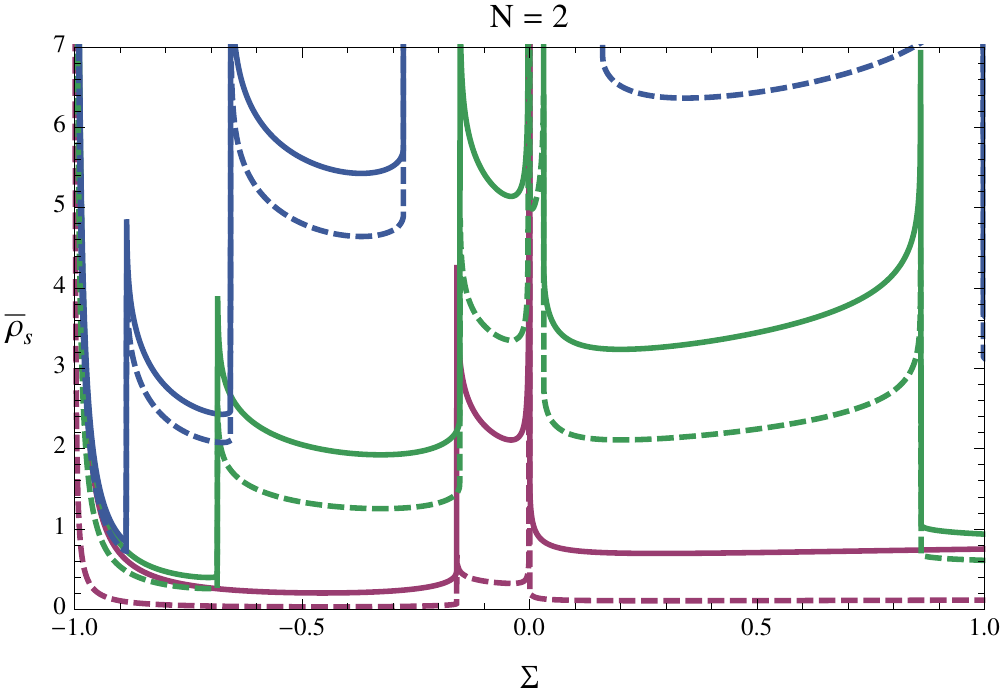}\qquad\includegraphics[scale=0.7]{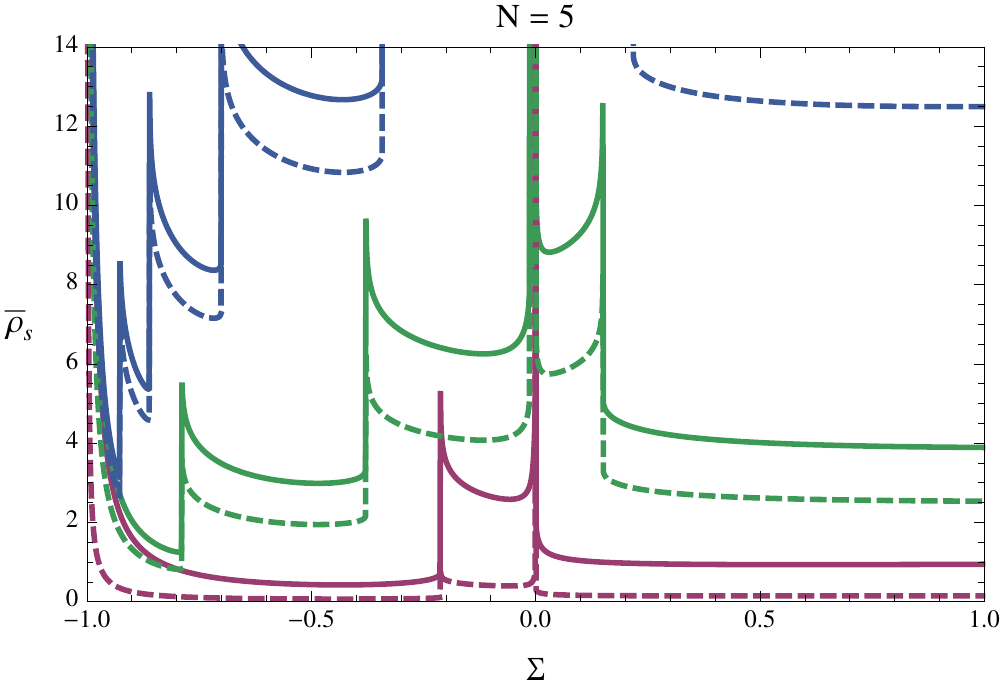}
\caption{
Each value of $\Sigma \equiv \bar{U}_{*}'(0)$ 
for which we have a spike in the above figure is the derivative at the origin 
of a well defined fixed point effective potential: thus every spike 
is the signature of a different universality class. 
Solid lines represents spike plots of LR models 
in dimension $d$ with power--law exponent $\sigma$, 
while dashed lines represent spike plots of 
SR models in dimension $D=D_{\rm eff}=2d/\sigma$.
The plot is for the case $N=1,2,5$ and $d=2$ for the cases 
$\sigma = 1.25,1.75,1.9$.}
\label{Fig1_app}
\end{figure}

We first study the flow Eq. \eqref{EffectivePotentialLPAprimeEquation_app} in the case $Z_{k}=1$, i.e. we set $\delta \eta = 0$. 
For comparison we report the analogous flow equation for the 
effective potential of the short--range (SR) model \cite{Delamotte07}:
\begin{equation}
\partial_t \bar{U}_{k}= -D\bar{U}_{k}(\bar{\rho})+(D-2+\eta_{SR})\bar{\rho}\,\bar{U}'_{k}(\bar{\rho})
+c_D (N-1)\frac{1-\frac{\eta_{SR}}{D+2}}{1+\bar{U}'_{k}(\bar{\rho})}+c_D \frac{1-\frac{\eta_{SR}}{D+2}}{1+\bar{U}'_{k}(\bar{\rho})+2\bar{\rho}\,\bar{U}''_{k}(\bar{\rho})}\,.
\label{EffectivePotentialLPAprimeEquationSR}
\end{equation}
Here we denote by $D$ the dimension of the SR model, while 
$d$ is the dimension of the lattice in which the LR model is defined. 
The key point of our analysis is that if we make the substitution
\begin{equation}
\label{DimensionaEquivalence_app}
D=D_{\rm eff}=\frac{2d}{\sigma}\,,
\end{equation}
in \eqref{EffectivePotentialLPAprimeEquationSR} we obtain again Eq.\,\eqref{EffectivePotentialLPAprimeEquation_app} with $\delta\eta=0$, 
apart for a factor $\frac{\sigma}{2}$ multiplying the scale derivative 
of the potential.
When we study the fixed point effective potential 
$\bar{U}_{*}(\bar{\rho})$ the scale derivative term in 
\eqref{EffectivePotentialLPAprimeEquation_app} vanishes
and there is no difference between 
\eqref{EffectivePotentialLPAprimeEquation_app} and 
\eqref{EffectivePotentialLPAprimeEquationSR} 
with $D=D_{\rm eff}=2d/\sigma$,  as it is
shown in Fig.\,1 of the main text.

We plot in Fig.\ref{Fig1_app} the results obtained for three $O(N)$ models: 
$N=1$ (Ising model), $N=2$ (XY model) and $N=5$
(similar plots can be drawn for any $N$). 
From now on we reabsorb the coefficients $c_d$ and $c_D$ 
in the definition of the field, following 
the same procedure described in \cite{Morris94}.
We now establish the mapping, valid within this approximation, 
between the LR correlation length exponent $\nu_{LR}(d,\sigma)$ and the
equivalent SR one $\nu_{SR}(D_{\rm eff})$ at the effective dimension.
This can be done following the procedure in \cite{Morris94} to evaluate
these exponents.
%
%We obtain that the LPA equations for the flow of the potential in the
%SR and LR case are:
%%
%\begin{subequations}
%\begin{equation}
%\begin{split}
%\label{EffectivePotentialLPAEquationLR}
%&\frac{\partial U_{k}}{\partial\log k}=d\bar{U}_{k}(\bar{\rho})-(d-\sigma)\bar{\rho}\,\bar{U}'_{k}(\bar{\rho})\\&-\frac{\sigma}{2}(N-1)\frac{1}{1+\bar{U}'_{k}(\bar{\rho})}-\frac{\sigma}{2}\frac{1}{1+\bar{U}'_{k}(\bar{\rho})+2\bar{\rho}\,\bar{U}''_{k}(\bar{\rho})}\,;
%\end{split}
%\end{equation}
%\begin{equation}
%\begin{split}
%\label{EffectivePotentialLPAEquationSR}
%&\frac{\partial U_{k}}{\partial\log k}=D\bar{U}_{k}(\bar{\rho})-(D-2)\bar{\rho}\,\bar{U}'_{k}(\bar{\rho})\\&-(N-1)\frac{1}{1+\bar{U}'_{k}(\bar{\rho})}-\frac{1}{1+\bar{U}'_{k}(\bar{\rho})+2\bar{\rho}\,\bar{U}''_{k}(\bar{\rho})}\,.
%\end{split}
%\end{equation}
%\end{subequations}
%
In order to calculate these exponents we have to 
write an eigenvalue equation for the stability of the perturbations 
around the scaling solution and then make the substitution 
$$\bar{U}_{k}(\bar{\rho})=\bar{U}_{*}(\bar{\rho})+k^{y}\bar{u}_k(\bar{\rho})$$
in Eqs.\,\eqref{EffectivePotentialLPAprimeEquation_app} and 
\eqref{EffectivePotentialLPAprimeEquationSR}.
The $y$s are the renormalization group eigenvalues and 
the correlation length critical exponent is determined by the relation $\nu^{-1}=y_{t}=\min\{y\}$. 
The eigenvalue equations for the LR and SR perturbation $\bar{u}_k(\bar{\rho})$ are, respectively, the following:
% and $\lambda$ are:
%
\begin{subequations}
\begin{equation}
\label{StabilityLPAEquationLR}
(d-y_{LR})\bar{u}_k(\bar{\rho})-(d-\sigma)\bar{\rho}\,\bar{u}_k'(\bar{\rho})+\frac{\sigma}{2}\frac{(N-1)\bar{u}_k'(\bar{\rho})}{(1+\bar{U}'_{*}(\bar{\rho}))^{2}}
-\frac{\sigma}{2}\frac{\bar{u}_k'(\bar{\rho})+2\bar{\rho}\,\bar{u}_k''(\bar{\rho})}{(1+\bar{U}'_{*}(\bar{\rho})+2\bar{\rho}\,\bar{U}''_{*}(\bar{\rho}))^{2}}=0\,,
\end{equation}
and
\begin{equation}
\label{StabilityLPAEquationSR}
(D-y_{SR})\bar{u}_k(\bar{\rho})-(D-2)\bar{\rho}\,\bar{u}_k'(\bar{\rho})+\frac{(N-1)\bar{u}_k'(\bar{\rho})}{(1+\bar{U}'_{*}(\bar{\rho}))^{2}}-\frac{\bar{u}_k'(\bar{\rho})+2\bar{\rho}\,\bar{u}_k''(\bar{\rho})}{(1+\bar{U}'_{*}(\bar{\rho})+2\bar{\rho}\,\bar{U}''_{*}(\bar{\rho}))^{2}}=0\,,
\end{equation}
\end{subequations}
where $\bar{U}_{*}(\bar{\rho})$ is the scaling solution and the boundary condition is given 
in \cite{Morris94}.
Evaluating the SR equation in dimension $D=D_{\rm eff}$ and multiplying 
both sides for $\sigma/2$ gives the result reported in the main text:
\begin{equation}
\label{NuRelationLPA_app}
\nu_{LR}(d,\sigma)=\frac{2}{\sigma}\nu_{SR}(D_{\rm eff})\,.
\end{equation}
Let us now consider the approximation in which the wavefunction
renormalization $Z_{k}$ is running but field independent and study 
Eq.\,\eqref{EffectivePotentialLPAprimeEquation_app} 
in the case $\delta\eta\neq 0$. Defining $Z_k$ as
\begin{equation}
\label{AnomalousDimensionFirstDefinition}
Z_{k}=\lim_{p\rightarrow 0}\frac{d}{dp^{\sigma}}\Gamma_{k}^{(2)}(p,-p)\,,
\end{equation}
leads to the following result: $$\delta\eta=0\,.$$
This is due to the peculiar properties of LR interactions, which lead to a non analytic term in the propagator. However when we calculate
the RG time derivative of the propagator it does not present any non--analytic, at our approximation level, thus the flow leaves the
$Z_{k}$ unaltered.

Thus in Eq.\,\eqref{EffectivePotentialLPAprimeEquation_app} we can just drop the $\delta \eta$ terms also in this case.
%
%The flow equation for the potential using LPA$'$ is different for
%SR and LR models. For the LR model one finds
%%
%\begin{subequations}
%\begin{equation}
%\begin{split}
%\label{EffectivePotentialLPAEquationLR2}
%&\frac{\partial U_{k}}{\partial\log k}=d\bar{U}_{k}(\bar{\rho})-(d-\sigma)\bar{\rho}\,\bar{U}'_{k}(\bar{\rho})\\&-\frac{\sigma}{2}(N-1)\frac{1}{1+\bar{U}'_{k}(\bar{\rho})}-\frac{\sigma}{2}\frac{1}{1+\bar{U}'_{k}(\bar{\rho})+2\bar{\rho}\,\bar{U}''_{k}(\bar{\rho})}\,,
%\end{split}
%\end{equation}
%%
%while for the SR model it is 
%%
%\begin{equation}
%\begin{split}
%\label{EffectivePotentialLPAEquationSR2}
%&\frac{\partial U_{k}}{\partial\log k}=D\bar{U}_{k}(\bar{\rho})-(D-2+\eta_{SR})\bar{\rho}\,\bar{U}'_{k}(\bar{\rho})\\&-(N-1)\frac{1-\frac{\eta_{SR}}{D+2}}{1+\bar{U}'_{k}(\bar{\rho})}-\frac{1-\frac{\eta_{SR}}{D+2}}{1+\bar{U}'_{k}(\bar{\rho})+2\bar{\rho}\,\bar{U}''_{k}(\bar{\rho})}\,,
%\end{split}
%\end{equation}
%\end{subequations}
%%
%where $\eta_{SR}$ satisfies LPA$'$ for SR models in fractional dimension,
%while $\delta\eta$ sticks to zero.
%
Proceeding in the same way as done in the case $Z_{k}=1$ (performing an additional rescaling of the field),
we obtain a new dimensional equivalence:
\begin{equation}
\label{EffectiveDimensionLPA$'$_app}
D_{\rm eff}'=\frac{[2-\eta_{SR}(D_{\rm eff}')]d}{\sigma}\,.
\end{equation}
At date the most precise evaluation of $\eta_{SR}$ in fractional dimension 
$D$ for the Ising model ($N=1$) is given in \cite{El14};
results for general $N$ are given in \cite{Codello13},
turning in rather good agreement with \cite{El14} for $N=1$ 
and with \cite{Holovatch93} for $N \ge 2$.
Using these results we can then evaluate $D_{\rm eff}'$ 
for various values of $N$ as shown in Fig.\,\ref{FigNuova}.
The fact that the $N\geq2$ curves reach two for $\sigma=2$ is due to the Mermin--Wagner theorem.
\begin{figure}
\centering
\includegraphics[scale=.9]{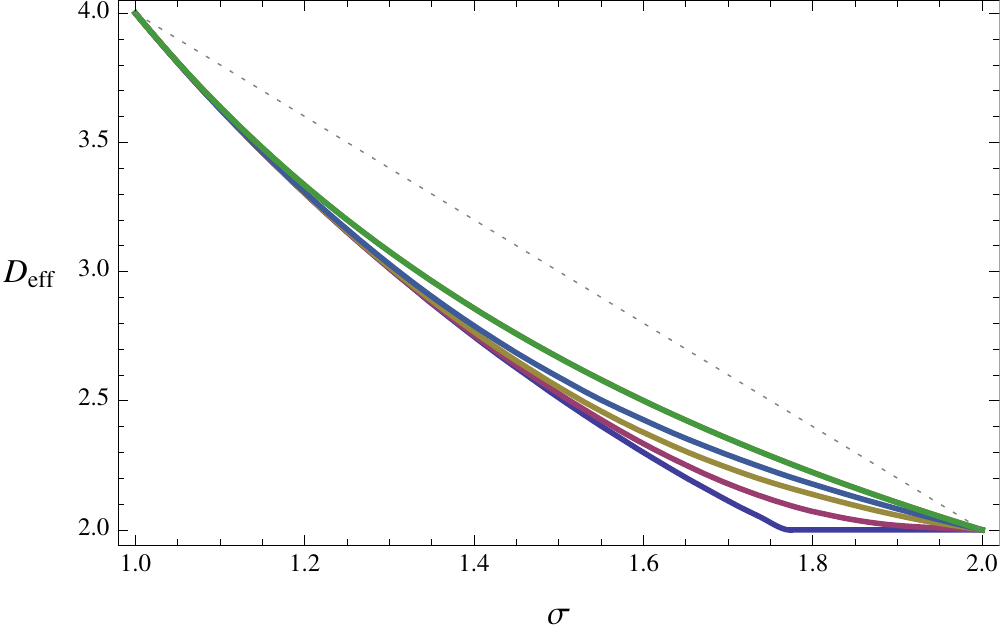}
\caption{
Different results for the effective dimension $D_{\rm eff}'$ of a 
LR model $O(N)$ model in $d=2$:Ä
(from left to right) $N = 1,2,3,5,100$ using the data from \cite{Codello13,Codello14}.
The $ N = 100 $ case already overlaps with $D_{\rm eff} = 2d/\sigma$ valid in the large--$N$ limit.
%the blue line is the effective dimension 
%in the LPA which is valid in the $N=\infty$ limit, the red line is 
%the effective dimension obtained in the LPA$'$ for the Ising model 
%using $\eta_{SR}$ of \cite{El14}, and the other lines between them are the 
%values of $D_{\rm eff}'$ using $\eta_{SR}$ of \cite{Codello13}. 
The gray dashed line is the proposal made in \cite{Blanchard13}.}
\label{FigNuova}
\end{figure}
Following the previous procedure to compute the correlation length critical exponent now leads to the following relation:
%Using Eq.\,\eqref{AnomalousDimensionFirstDefinition} in LPA$'$ we also 
%obtain for the critical exponent $\nu_{LR}$ the following relation:
%
%THIS EQUATION IS ALREADY PRESENT IN THE MAIN TEXT
\begin{equation}
\label{NuRelationLPA2_app}
\nu_{LR}(d,\sigma)=\frac{2-\eta_{SR}(D_{\rm eff}')}{\sigma}\, \nu_{SR}(D_{\rm eff}')\,.
\end{equation}
The comparison of the exponent $y_{t}=1/\nu_{LR}$ for the LR Ising 
in $d=2$ obtained using (\ref{NuRelationLPA2_app}) and Monte Carlo results 
from \cite{Luijten02} and \cite{Parisi14} 
is plotted in the inset of the Fig.\,2 of the main text.
Here we report other useful comment:
the agreement is rather good for $\sigma \lesssim 1.75$, 
while for $\sigma \gtrsim 1.75$ the agreement 
becomes worst: this is due to the fact that in the SR case at this 
approximation level $\eta_{SR}(D=2)=0.233$ 
and then according to Sak one would have 
$\sigma_*=1.767$ which is not the exact value $\frac{7}{4}$ provided by Sak. 
Therefore, even if the result $\delta\eta=0$ is in agreement 
with \cite{Sak73} our prediction for $\nu_{LR}$ (blue line) 
has its own error due to the approximation of field independent
wavefunction renormalization 
(e.g., the value of $\nu_{SR}(d=2)=1.05$, 
is not the exact one $\nu_{SR}(d=2)=1$). 
This is confirmed from the fact that using the numerically exact 
values of $\eta_{SR}$ in the equation \eqref{NuRelationLPA2_app} (showed as the top pink line of the inset) agreement 
with MC results greatly improves between $\sigma=\frac{7}{4}$ and $\sigma=2$.
\section{Competing interactions}\label{Appendix_B}
According to Eq. \eqref{EffectiveDimensionLPA$'$_app} 
the value of the decay exponent for which we recover SR 
behaviour is $\sigma_*=2-\eta_{SR}$, in agreement with 
Sak's result \cite{Sak73}. However in this case we are not able to 
investigate the behaviour of the system above this threshold 
since we are not including any $p^{2}$ term in our 
ansatz \eqref{EffectiveAction_app}. On the other hand 
it is crucial to verify whether the system is actually 
recovering all its SR features above $\sigma_*$ or 
if it is still holding some LR properties. 
In order to pursue this investigation we enlarge our 
theory space. Our new ansatz is
\begin{equation}
\label{EffectiveActionLPA$''$}
\begin{split}
\Gamma_k[\phi]=&\int d^{d}x \Bigl{\{ } Z_{\sigma}\partial_{\mu}^{\frac{\sigma}{2}}\phi_{i}\partial_{\mu}^{\frac{\sigma}{2}}\phi_{i}
+Z_{2}\partial_{\mu}\phi_{i}\partial_{\mu}\phi_{i}+U_{k}(\rho)\Bigr{ \} }\,.
\end{split}
\end{equation}
It is quite straightforward to follow the same procedure given 
in the previous section using the following generalized Litim cutoff,
\begin{equation}
\label{Generalized_Litim_Cutoff}
R_{k}(q)\!=\!Z_{\sigma}(k^{\sigma}-q^{\sigma})\theta(k^{\sigma}-q^{\sigma})
\!+\!Z_{2}(k^{2}-q^{2})\theta(k^{2}-q^{2})\,,
\end{equation}
this cutoff has the desired property to not choose any term as the relevant one, it acts on both terms, making us sure to be valid in the whole $\sigma$ 
range. The choice (\ref{Generalized_Litim_Cutoff}) turns to be 
the most simple, since it always influences the dominant term, 
while only adding an irrelevant modification to the other, 
yet it drastically simplifies the calculation.\\
We proceed deriving the flow equation for all the quantities in latter definition, once again we have,
\begin{align}
\label{Wave_Function_Flow_Definition}
\partial_t Z_{2}&=\frac{1}{2}\lim_{p\rightarrow 0}\frac{d^{2}}{dp^{2}} \partial_t \Gamma^{(2)}_{k}(p,-p)\\
\partial_t Z_{\sigma}&=\lim_{p\rightarrow 0}\frac{d}{dp^{\sigma}}\partial_t \Gamma^{(2)}_{k}(p,-p)\,,
\end{align}
while the flow for the potential derives from the flow of the effective action evaluated at constant fields. These equations were obtained starting from the usual Wetterich 
Eq. \cite{Wetterich93}, 
with the ansatz \eqref{EffectiveActionLPA$''$}. 
The cutoff function is shown in Eq. \eqref{Generalized_Litim_Cutoff}. 
We firstly derived the equations for dimensional quantities,
\begin{subequations}
\begin{equation}
\label{Dimensional_Z_Sigma_Flow_app}
\partial_t Z_{\sigma}=0\,,
\end{equation}
\begin{equation}
\label{Dimensional_Z_2_Flow_app}
\partial_t Z_{2}=
\!-\!\frac{\rho_{0}\,U''_{k}(\rho_{0})^{2}\,(\sigma Z_{\sigma}k^{\sigma}\!\!+\!2Z_{2})^{2}k^{d+2}}{(Z_{\sigma}k^{\sigma}\!+\!Z_{2}k^{2})^{2}(Z_{\sigma}k^{\sigma}\!+\!Z_{2}k^{2}\!+\!2\bar{\rho}_{0}U''_{k}(\rho_{0}))^{2}}\,,
\end{equation}
\begin{equation}
\label{Dimensional_Potential_Flow_app}
\partial_t U_{k}(\rho)=%\left(Z_{2}k^{2}-\frac{\partial_t Z_{2}}{d+2}+\frac{\sigma}{2}Z_{\sigma}\right)\\
\;\frac{Z_{2}k^{2}-\frac{\partial_t Z_{2}}{d+2}+\frac{\sigma}{2}Z_{\sigma}}{Z_{\sigma}k^{\sigma}+Z_{2}k^{2}+U'_{k}(\rho)+2\rho U''_{k}(\rho)}
+(N-1)\frac{Z_{2}k^{2}-\frac{\partial_t Z_{2}}{d+2}+\frac{\sigma}{2}Z_{\sigma}}{Z_{\sigma}k^{\sigma}+Z_{2}k^{2}+U'_{k}(\rho)}\,.
\end{equation}
\end{subequations}
To further proceed we need to 
choose the dimension of the field with the constraint 
that the effective action must be dimensionless.
To properly define the dimensionless couplings, we have two natural choices:
the first one is the one we did in previous section 
to make the $Z_{\sigma}$ coupling dimensionless and  
absorb it into the field -- we refer to this choice as to 
{\em LR-dimensions}. On the other hand in this case we could also follow
the usual way for $O(N)$ models defining the 
field dimension to make $Z_{2}$ dimensionless and then absorbing
it in the field. 
This will lead to the definition of a LR coupling 
$J_{\sigma}=\frac{Z_{\sigma}}{Z_{2}}$ ({\em SR-dimensions}).\\
The two possible choices are summarized in the following table:
\begin{center}
\begin{tabular}{ccc}
{\em Quantity \quad \/} & {\em SR--dimensions \quad \/} & {\em LR--dimensions\/}\\
$q$ & $k\bar{q}$ & $k\bar{q}$ \\
$\rho$ & $k^{d-2} Z_{2}^{-1} \bar{\rho}$ & $k^{d-\sigma} Z_{\sigma}^{-1} \bar{\rho}$ \\
$U(\rho)$ & $k^{d}\bar{U}(\bar{\rho})$ &  $k^{d}\bar{U}(\bar{\rho})$\\
$Z_{2}$ & $\bar{Z}_{2}$ &  $k^{\sigma-2}\bar{Z}_{2}$\\
$Z_{\sigma}$ & $k^{2-\sigma}\bar{Z}_{\sigma}$ &  $\bar{Z}_{\sigma}$\\
\end{tabular}
\end{center}
Physical results should be the same in both cases. 
If we choose SR-dimensions we find three equations: 
one for the potential, one for the
LR coupling and one for the anomalous dimension. 
These three equations reproduce the usual $O(N)$ models equations 
in the limiting case $J_{\sigma}\rightarrow 0$. 
On the other hand when we use LR-dimensions 
we have only two equations (since we do not have any anomalous dimension)
and we may define a SR coupling $J_{2}=Z_{2}/Z_{\sigma}$: 
when $J_{2}$ runs to zero we recover the equations
obtained for the pure LR approximation.

We conclude then that our last includes the results 
obtained in previous ones and extends them in the whole $\sigma$ range. 
SR-dimensions prove better to investigate the boundary 
$\sigma\simeq\sigma_*$, due to the fact that $Z_{\sigma}$ 
is always constant during the flow, while $Z_{2}$ is diverging 
in the case of dominant SR interactions and must be absorbed in the field.
\subsection*{SR--Dimensions}
We are going to investigate the region 
$\sigma>\sigma_*$, where we believe the $p^{2}$ 
term to be dominant, so we choose the SR-dimensions 
in order to be able to recover exactly the SR case.
We define our anomalous dimension as
\begin{equation}
\eta_{2}=-\frac{1}{Z_{2}} \partial_t Z_{2} \,,%\frac{d Z_{2}}{d\log k}\,,
\end{equation}
(following the usual SR analysis \cite{Delamotte07}), 
but in addition one gets the renormalized LR coupling defined as
\begin{equation}
J_{\sigma}=\frac{Z_{\sigma}}{Z_{2}}\,.
\end{equation}
The flow equations for the renormalized dimensionless couplings are
\begin{subequations}
\begin{equation}
\label{J_Flow}
\partial_t \bar{J}_{\sigma} = (\sigma-2)\bar{J}_{\sigma}+\eta_{2}\bar{J}_{\sigma}\,,
\end{equation}
\begin{equation}
\eta_{2}=\frac{(2+\sigma\bar{J}_{\sigma})^{2}\bar{\rho}_0\bar{U}''_{k}(\bar{\rho}_0)^{2}}{(1+\bar{J}_{\sigma})^{2}(1+\bar{J}_{\sigma}+2\bar{\rho}_0\bar{U}''_{k}(\bar{\rho}_0))^{2}}\,,
\end{equation}
\begin{equation}
\begin{split}
\label{PotentialFlow_app}
\partial_t \bar{U}_{k}(\bar{\rho})=&-d\bar{U}_{k}(\bar{\rho})+(d-2+\eta_{2})\bar{\rho}\,\bar{U}'_{k}(\bar{\rho})
%&\qquad\qquad-\left(1-\frac{\eta_{2}}{d+2}+\frac{\sigma}{2}\bar{J}_{\sigma}\right) \\
+(N-1)\frac{1-\frac{\eta_{2}}{d+2}+\frac{\sigma}{2}\bar{J}_{\sigma}}{1+\bar{J}_{\sigma}+\bar{U}'_{k}(\bar{\rho})}\\
&+\frac{1-\frac{\eta_{2}}{d+2}+\frac{\sigma}{2}\bar{J}_{\sigma}}{1+\bar{J}_{\sigma}+\bar{U}'_{k}(\bar{\rho})+2\bar{\rho}\,\bar{U}''_{k}(\bar{\rho})}\,.
\end{split}
\end{equation}
\end{subequations}
Looking at Eq. \eqref{J_Flow} we see that there 
are only two possibilities for the r.h.s. 
to vanish and for $\bar{J}_{\sigma}$ to attain 
some fixed point value $\bar{J}_{\sigma}^{*}$. \\
The first possibility is $\bar{J}_{\sigma}^{*}=0$ and we are in the SR case, 
the second is $\eta_{2}=2-\sigma$ which is a characteristic of the LR fixed point, a least at this approximation level.
\begin{figure}
\centering
\includegraphics[scale=1]{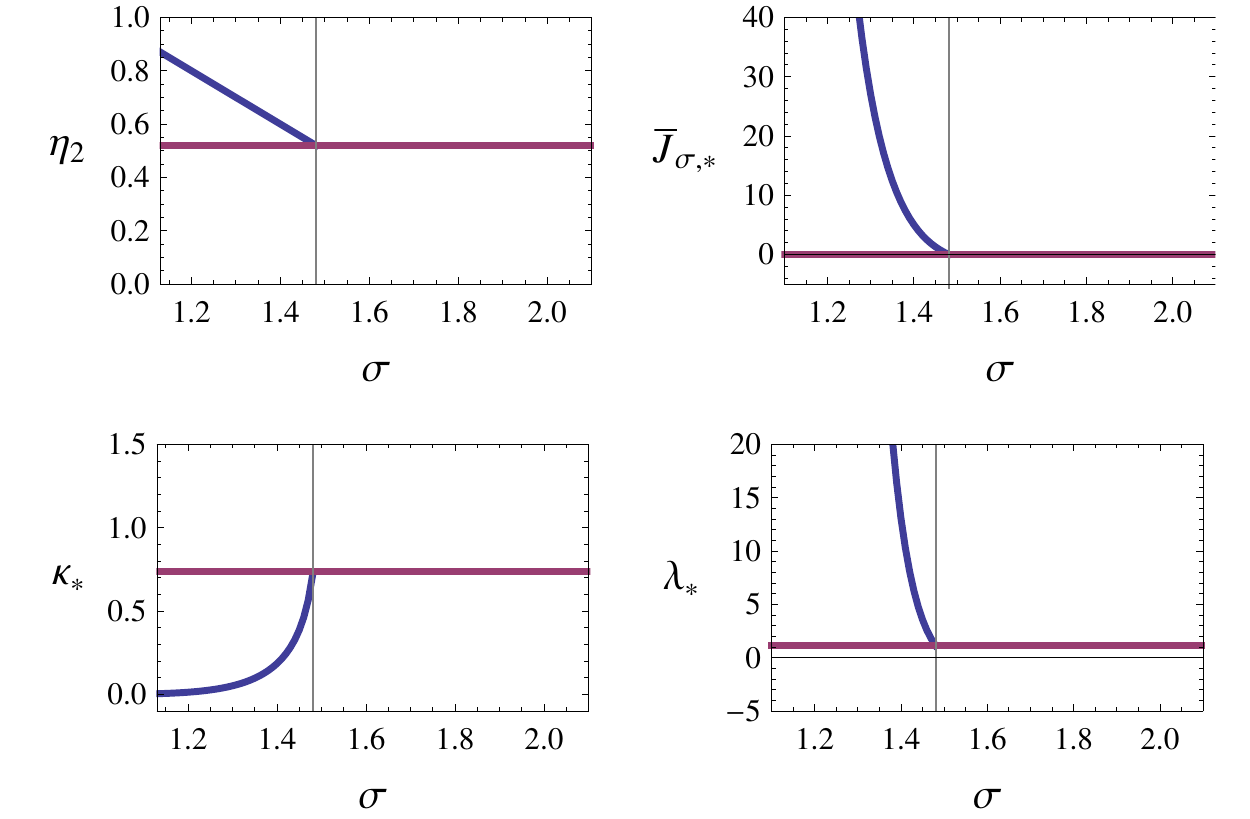}
\caption{Anomalous dimension $\eta_2$ and fixed point values $J_{\sigma,*}$,$\kappa_*$,$\lambda_*$ in the truncation  considered  in the text. For $\sigma> \sigma_* \equiv 2-\eta_{SR}$ only the fixed point (red line) is present characterized by $\eta_2=\eta_{SR}$ and $J_{\sigma,*}=0$. At  $\sigma=\sigma_*$ the  LR fixed point (blue lines) branches from the SR fixed point and then controls the critical behaviour for every $\sigma<\sigma_*$. Thus even in the case of both SR and LR terms in the propagator the anomalous dimension as a function of $\sigma$ is thus LR ($\eta_2=2-\sigma$) for $\sigma<\sigma_*$ and SR for $\sigma>\sigma_*$.}  
\label{FIGURE_APP3}
\end{figure}
This shows that we have no necessity to change the 
field dimension to study the case of a dominant LR term, 
since the $p^{2}$ term is still present in the LR fixed point.\\
In order to check these properties we turn to the approximation where we
expand the potential around its minimum:
\begin{equation}
\label{Potential_Expansion_Around_Minimum_LPA$'$}
\bar{U}_{k}(\bar{\rho})=\frac{1}{2}\lambda_{k}(\bar{\rho}-\kappa_{k})^{2}\,.
\end{equation}
Projecting the flow equation for the potential we can 
get the beta functions of these two couplings which, 
together with the flow equation for $\bar{J}_{\sigma}$, 
form a closed set:
\begin{subequations}
\begin{equation}
\partial_{t}\bar{J}_{\sigma}=(\sigma-2)\bar{J}_{\sigma}+\eta_{2}\bar{J}_{\sigma}\,,
\end{equation}
\begin{equation}
\eta_{2}=\frac{(2+\sigma\bar{J}_{\sigma})^{2}\kappa_{k}\lambda_{k}^{2}}{(1+\bar{J}_{\sigma})^{2}(1+\bar{J}_{\sigma}+2\kappa_{k}\lambda_{k})^{2}}\,,
\end{equation}
\begin{equation}
\label{opportuna}
\partial_{t}\kappa_{k} =  \;   -(d-2+\eta_2 )\kappa_k + 3\frac{1-\frac{\eta_2 }{d+2}+\frac{\sigma}{2}\bar{J}_{\sigma}}{(1+\bar{J}_{\sigma}+2 \kappa_k  \lambda_k)^2}+(N-1)\frac{1-\frac{\eta_2}{d+2}+\frac{ \sigma}{2}\bar{J}_{\sigma}}{(1+\bar{J}_{\sigma})^2}\,,
\end{equation}
\begin{equation}
\partial_{t} \lambda_{k} = (d-4+2\eta_{2})\lambda_{k} + 18 \lambda_k\frac{1-\frac{\eta_2 }{d+2}+\frac{\sigma}{2}\bar{J}_{\sigma}}{(1+\bar{J}_{\sigma}+2 \kappa_k  \lambda_k)^3}
+ 2 \lambda_k(N-1)\frac{1-\frac{\eta_2}{d+2}+\frac{ \sigma}{2}\bar{J}_{\sigma}}{(1+\bar{J}_{\sigma})^3}\,.
\end{equation}
\end{subequations}
As discussed SR-dimensions are well suited to study 
the case $\sigma<\sigma_*$, since $Z_{2}$ is well defined 
in this case and is not brought to zero by the presence of a dominant LR term. 
The results for the couplings and the anomalous dimension at the fixed point
is shown in Fig.\ref{FIGURE_APP3}. We see that the anomalous dimension $\eta_{2}$ follows naturally the Sak's behaviour with no possible LR fixed point solution for $\sigma>\sigma_*$, however at $\sigma=\sigma_*$ the LR fixed point (blue lines) appears and the value of the coupling in that point is shown. 
It is possible to see that the $\bar{J}_{\sigma,*}$  grows very fast when we approach $\sigma=d/2$ (which is $1$  in this case since we are plotting $2$ dimensional results). This coupling is actually diverging  at $\sigma=1$ since at the point the LR fixed point merges with the Gaussian LR fixed point, which can be suitably described only in LR-dimensions, since it has no SR term in its propagator.
This has been verified for different values of 
$\sigma>\sigma_*$, at different (also non-integer) 
dimensions and for various $O(N)$ models. 
It is also possible to show that 
the truncation of the potentials despite changing 
the value of any quantity at the fixed point does 
not modify the qualitative behaviour of
the system nor the existence of the threshold $\sigma_*$.\
\begin{figure}
\centering
\includegraphics[scale=1]{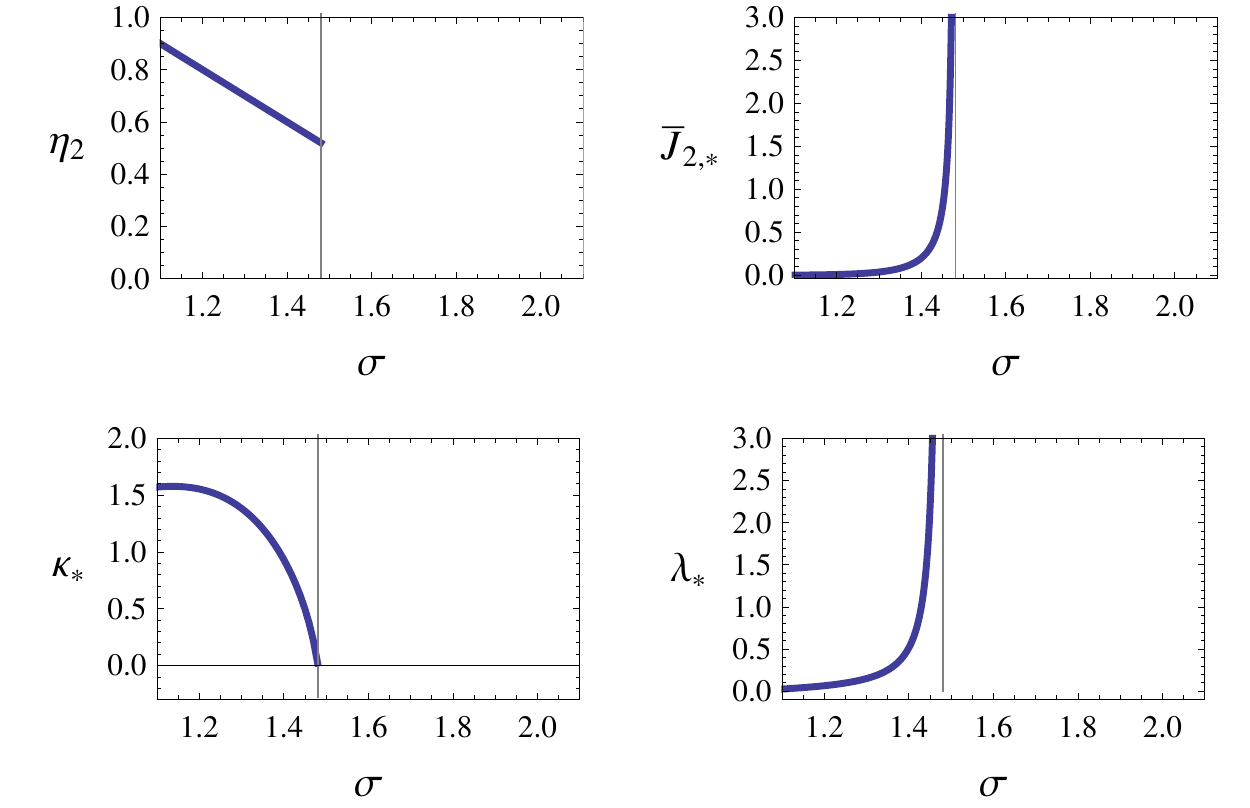}
\caption{Anomalous dimension $\eta_2$ and fixed point values $\bar{J}_{\sigma,*},\kappa_*,\lambda_*$ in the truncation \eqref{EffectiveActionLPA$''$} with LR--dimensions.
With this dimensional choice we are able to describe only the $\sigma < \sigma_*$ region,
since in the other region the coupling $\bar{J}_{2,k}$ is divergent, as can be understood from the $\bar{J}_{2,k}$ plot.
Also in this case, the anomalous dimension as a function of $\sigma$ is LR ($\eta_2=2-\sigma$) for $\sigma < \sigma_*$.
}
\label{FIGURE_APP4}
\end{figure}
\subsection*{LR--Dimensions}
Here for the sake of completeness we also report the results obtained using 
LR dimensionless variables. In this case we renormalize the field using the wave function $Z_{\sigma}$ and we define the SR coupling $\bar{J}_{2}=\bar{Z}_{2}/Z_{\sigma}$
\begin{subequations}
\begin{equation}
\label{Dimensionless_Z_Sigma_Flow_app}
\partial_t Z_{\sigma}=0\,,
\end{equation}
\begin{equation}
\label{J_2_Flow_app}
\partial_t \bar{J}_{2}=(2-\sigma)\bar{J}_{2}-\frac{\bar{\rho}_0\,\bar{U}''_{k}(\bar{\rho}_0)^{2}(\sigma+2\bar{J}_{2})^{2}}{(1+\bar{J}_{2})^{2}(1+\bar{J}_{2}+2\kappa_{k}\bar{U}''_{k}(\bar{\rho}_0))^{2}}\,,
\end{equation}
\begin{equation}
\label{Dimensionless_Potential_Flow_app}
\begin{split}
\partial_t U_{k}(\rho)=&-d\bar{U}_{k}(\bar{\rho})+(d-\sigma)\bar{\rho}\,\bar{U}'_{k}(\bar{\rho})+\frac{\bar{J}_{2}-\frac{(2-\sigma)\bar{J}_{2}+\partial_t \bar{J}_{2}}{d+2}+\frac{\sigma}{2}}{1+\bar{J}_{2}+\bar{U}'_{k}(\bar{\rho})+2\bar{\rho} \,\bar{U}''_{k}(\bar{\rho})}\\
&+(N-1)\frac{\bar{J}_{2}-\frac{(2-\sigma)\bar{J}_{2}+\partial_t \bar{J}_{2}}{d+2}+\frac{\sigma}{2}}{1+\bar{J}_{2}+\bar{U}'_{k}(\bar{\rho})}\,.
\end{split}
\end{equation}
\end{subequations}
These equations in the $\bar{J}_{2}\rightarrow 0$ limit reproduce the 
results obtained for previous the approximations. 
Thus this approximation reproduces, as expected, 
all the previous results in the range $\sigma<\sigma_*$, 
but also gives information on their validity, 
comparing latter equation with \eqref{EffectivePotentialLPAprimeEquation_app} 
we see that they are equal up to a term of order $\bar{J}_{2}$. 
In fact if we use LR-dimensions we find coherently that
 $\bar{J}_{2}^{*}\neq 0$ is very small for all  $\sigma<\sigma_*$  but in the region $\sigma\simeq\sigma_*$, when as we expected the effective dimension relations are spoiled (this is shown in Fig.\ref{FIGURE_APP4}).
\begin{figure}
\centering
\includegraphics[scale=1]{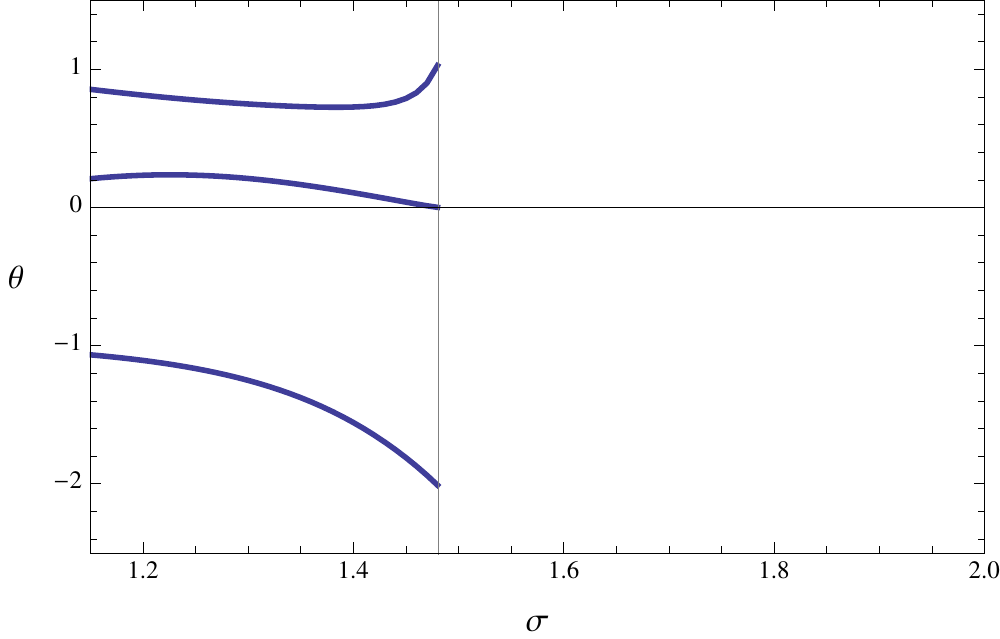}
\caption{Eigenvalues ($\theta$) of the RG stability matrix as a function of $\sigma$ for the LR (blue lines) fixed points in the case of LR-dimensions. In this case we are not able to describe only the LR fixed point present when $\sigma<\sigma_*$ and,since RG eigenvalues are universal quantities, they agree with SR-dimensions ones.}
\label{FIGURE_APP5}
\end{figure}

We can then repeat the previous analysis. We have in this case one, 
very important, difference: 
we are renormalizing the field with the $Z_{\sigma}$ wave function. 
This is consistent in the range $\sigma<\sigma_*$ where 
the LR interaction, while in the case of a dominant $p^{2}$ interaction 
($\sigma>\sigma_*$) we expect the $Z_{2}$ wave function to be diverging, 
this divergence is not absorbed in the field as an 
anomalous dimension and cannot be balanced by $Z_{\sigma}$ 
which we know to be constant at this approximation level. 
Thus this divergence will still be present in our flow and this choice for the dimensionless coupling is not suited in the case $\sigma>\sigma_*$ (Fig. \ref{FIGURE_APP4}).

We then investigate the case $\sigma<\sigma_*$ where our variable 
are well defined, as usual we refer to the very simple truncation 
shown in Eq. \eqref{Potential_Expansion_Around_Minimum_LPA$'$} and 
we use the renormalization time $t=\log (k/k_0)$. The flow equations are:
\begin{subequations}
\begin{equation}
\begin{split}
\partial_{t}\bar{J}_{2}=(2-\sigma)\bar{J}_{2}+\frac{\kappa_{k}\lambda_{k}^{2}(\sigma+2\bar{J}_{2})^{2}}{(1+\bar{J}_{2})^{2}(1+\bar{J}_{2}+2\kappa_{k}\lambda_{k})^{2}}\,,
\end{split}
\end{equation}
\begin{equation}
\begin{split}\label{opportunabis}
\partial_{t}\kappa_{k} = & \;   -(d-\sigma)\kappa_k + 3\frac{\bar{J}_{2}-\frac{(2-\sigma)\bar{J}_{2}+\partial_t \bar{J}_{2}}{d+2}+\frac{\sigma}{2}}{(1+\bar{J}_{2}+2 \kappa_k  \lambda_k)^2}\\
&+ (N-1)\frac{\bar{J}_{2}-\frac{(2-\sigma)\bar{J}_{2}+\partial_t \bar{J}_{2}}{d+2}+\frac{\sigma}{2}}{(1+\bar{J}_{2})^2}\,,
\end{split}
\end{equation}
\begin{equation}
\begin{split}\label{opportunabis}
\partial_{t}\lambda_{k} = & \;   (d-2\sigma) \lambda_k +18 \lambda_k \frac{\bar{J}_{2}-\frac{(2-\sigma)\bar{J}_{2}+\partial_t \bar{J}_{2}}{d+2}+\frac{\sigma}{2}}{(1+\bar{J}_{2}+2 \kappa_k  \lambda_k)^3}\\
&+2 \lambda_k (N-1)\frac{\bar{J}_{2}-\frac{(2-\sigma)\bar{J}_{2}+\partial_t \bar{J}_{2}}{d+2}+\frac{\sigma}{2}}{(1+\bar{J}_{2})^3}\,.
\end{split}
\end{equation}
\end{subequations}
The fixed points solutions of the couplings is reported in Fig.\,\ref{FIGURE_APP4}, where we see that only the LR fixed points solution are present and then only the region $\sigma<\sigma_*$ is investigated. In fact, as well as previous couplings showed a diverging $\bar{J}_{\sigma,*}$ for $\sigma \to 1$, these couplings show a divergence in the limit $\sigma \to \sigma_*$ where the LR term in the propagator is vanishing.
However the results for the critical exponents in the region where both the couplings sets are defined are in perfect agreement between themselves, as it should be and as it is shown in Fig.\ref{FIGURE_APP5}.}


\begin{thebibliography}{XX}
\bibitem{Dauxois10} 
{\em Long-Range Interacting Systems: Lecture
Notes of Les Houches Summer School,} T. Dauxois, S. Ruffo, and L. F. 
Cugliandolo eds. (Oxford, Oxford University Press, 2010).

\bibitem{Dyson69}
F. J. Dyson, Comm. Math. Phys. {\bf 12}, 91 (1969).

\bibitem{Thouless69}
D. J. Thouless, Phys. Rev. {\bf 187}, 732 (1969).

\bibitem{Anderson70}
P. W. Anderson, G. Yuval, and D. R. Hamann,  
Phys. Rev. B {\bf 1}, 4464 (1970). 

\bibitem{Campa09}
A. Campa, T. Dauxois, and S. Ruffo,
Phys. Rep. {\bf 480}, 57 (2009).

\bibitem{Fisher72} 
M. E. Fisher, S. K. Ma, and B. G. Nickel, Phys. Rev. Lett. {\bf 29}, 14 (1972). 

\bibitem{Sak73}
J. Sak, Phys. Rev. B {\bf 8}, 281 (1973). 

\bibitem{Cardy81}
J. L. Cardy, J. Phys. A {\bf 14}, 1407 (1981).

\bibitem{Froelich82}
J. Fr\"olich and T. Spencer, Comm. Math. Phys. {\bf 84}, 87 (1982).

\bibitem{Luijten01}
E. Luijten and H. Messingfeld, 
Phys. Rev. Lett. {\bf 86}, 5305 (2001).

\bibitem{Luijten97}
E. Luijten, {\em Interaction Range, 
universality and the Upper Critical Dimension}, Ph. D. Thesis (1997).

\bibitem{Luijten95} 
E. Luijten and H. W. J. Bl\"ote, Int. J. Mod. Phys. C {\bf 6}, 359 (1995).

\bibitem{Luijten02}
E. Luijten and H. W. J. Bl\"ote, Phys. Rev. Lett. {\bf 89}, 025703 (2002).

\bibitem{Picco12}
M. Picco, \verb|arXiv:1207.1018|

\bibitem{Blanchard13}
T. Blanchard, M. Picco, and M. A. Rajapbour, 
Europhys. Lett. {\bf 101}, 56003 (2013).

\bibitem{Grassberger13}
P. Grassberger, \verb|arXiv:1305.5940|

\bibitem{Parisi14}
M. C. Angelini, G. Parisi, and F. Ricci-Tersenghi, 
Phys. Rev. E { \bf 89}, 062120 (2014).

\bibitem{Berges02}
J. Berges, N. Tetradis, and C. Wetterich, 
Phys. Rep. {\bf 363}, 223 (2002).

\bibitem{Delamotte07}
B. Delamotte, in {\em Order, disorder and criticality: 
advanced problems of phase transition theory}, 
Yu. Holovatch ed. (Singapore, World Scientific, 2007)
[\verb|arXiv:cond-mat/0702365|].

\bibitem{Codello13}
A. Codello and G. D'Odorico,
Phys. Rev. Lett. {\bf 110}, 141601 (2013).

\bibitem{Codello14}
A. Codello, N. Defenu, and G. D'Odorico,
\verb|arXiv:1410.3308|

\bibitem{Wetterich93}
C. Wetterich, Phys. Lett. B {\bf 301}, 90 (1993).

\bibitem{Codello12}
A. Codello, J. Phys. A {\bf 45}, 465006 (2012).

\bibitem{Morris94}
T. R. Morris, Phys. Lett. B {\bf 334}, 355 (1994).

\bibitem{Joyce}
G. S. Joyce, in {\em Phase Transitions and Critical Phenomena} vol. {\bf 2}, 
C. Domb and M. S. Green eds., pp. 375-442 
(London, Academic Press, 1972).

\bibitem{Tetradis94}
N. Tetradis and C. Wetterich, Nucl. Phys. B {\bf 422}, 541 (1994).

\bibitem{Stanley68}
H. E. Stanley, Phys. Rev. {\bf 176}, 718 (1968). 

\bibitem{Young12}
R. A. Ba\~nos, L. A. Fernandez, V. Martin-Mayor, and A. P. Young,
Phys. Rev. B {\bf 86}, 134416 (2012). 

\bibitem{Katz77}
S. L. Katz, M. Droz, and J. D. Gunton, Phys. Rev. B 
{\bf 15}, 1597 (1977).

\bibitem{Holovatch93}
Yu. Holovatch, Int. J. Mod. Phys. A {\bf 8}, 5329 (1993).

\bibitem{El14}
S. El-Showk, M. Paulos, D. Poland, S. Rychkov, D. Simmons-Dun, and A. Vichi, 
Phys. Rev. Lett. {\bf 112}, 141601 (2014).

\bibitem{Tissier14}
M. Baczyk, M. Tissier, G. Tarjus, and Y. Sakamoto
Phys. Rev. B { \bf 88}, 014204 (2013).

\bibitem{Balog14}
I. Balog, G. Tarjus, and M. Tissier
J. Stat. Mech. P10017 (2014).

\bibitem{Parisi14_2}
E. Brezin, G. Parisi, and F. Ricci-Tersenghi
arXiv:1407.3358.

\end{thebibliography}
\end{document}